\documentclass[fleqn,usenatbib]{mnras}
\usepackage{newtxtext,newtxmath}
\usepackage[T1]{fontenc}
\DeclareRobustCommand{\VAN}[3]{#2}
\let\VANthebibliography\thebibliography
\def\thebibliography{\DeclareRobustCommand{\VAN}[3]{##3}\VANthebibliography}

\usepackage{graphicx}	
\usepackage{amsmath}	
\usepackage{longtable}
\usepackage{adjustbox}
\usepackage{multirow}
\usepackage[version=4,arrows=pgf-filled,
mathfontname=mathsf]{mhchem} 
\newcommand{\HII}{H~{\sc ii}}

\title[Tracing Molecular Clumps Evolution with ML]{Identifying Evolutionary Stages of Molecular Clumps through Unsupervised and Supervised Machine Learning}
\author[K. V. Plakitina et al.]{
K. V. Plakitina$^{1}$,\thanks{E-mail: plakitina.kv@inasan.ru}
M. S. Kirsanova$^{1}$,
A. B. Ostrovskii$^{2}$,
A. D. Gimalieva$^{2}$,
S. V. Salii$^{2}$,
A.~V.~Meshcheryakov $^{3}$
\\
$^{1}$Institute of Astronomy of the Russian Academy of Sciences, 48 Pyatnitskaya Str., Moscow, Russia\\
$^{2}$Ural Federal University named after the first President of Russia B.N.Yeltsin, 19 Mira Str., Ekaterinburg, Russia\\
$^{3}$ Space Research Institute of the Russian Academy of Sciences, Profsoyuznaya Str. 84/32, Moscow, Russia\\
}

\date{Accepted XXX. Received YYY; in original form ZZZ}

\pubyear{\the\year{}}

\begin{document}
\label{firstpage}
\pagerange{\pageref{firstpage}--\pageref{lastpage}}
\maketitle

\begin{abstract}

The evolutionary classification of molecular clumps, crucial for understanding star formation, is commonly based on human-assigned categories derived from infrared (IR) emission and well-established morphological criteria. However, due to ambiguous signatures, distance uncertainties or heavily obscured IR emission, a significant fraction of sources often remains unclassified. This work demonstrates the capability of machine learning (ML) as a complementary, data-driven approach to automate the identification and classification of these clumps using data from the MALT90 survey, complemented by {\it Spitzer} IR photometry. We applied unsupervised clustering with HDBSCAN on molecular line intensities, revealing distinct groupings that correspond to evolutionary stages. Using only five molecular lines (HCO$^+$, HNC, N$_2$H$^+$, HCN, C$_2$H), we identified stable clusters of protostars and regions without active star formation, driven primarily by C$_2$H and N$_2$H$^+$ emission. Incorporating H$^{13}$CO$^+$ gave rise to a distinct UV-dominant cluster, tracing more evolved regions. Infrared properties appeared as non-significant features implying that envelopes of clumps with different masses are similar in their global infrared characteristics. We then employed supervised learning to classify clumps with previously uncertain categories and provided classifications for 522 objects, predominantly as regions without active star formation. Our results show that ML techniques can effectively uncover intrinsic evolutionary structures in complex astrochemical data and assign categories 
to uncertain sources, providing a powerful, data-driven complement to traditional methods.

\end{abstract}




\section{Introduction}

Modern astronomy is undergoing a transformation driven by the rapid growth of large-scale observational surveys. Advances in telescope technology have led to the collection of vast amounts of spectral information, particularly from molecular line observations. These data sets cover different stages of star formation, from  infrared dark clouds (IRDC's) and protostars to evolved \HII{} regions \citep{2009A&A...504..415S, 2021A&A...651A..85B, 2010PASP..122..314M,  2013wise.rept....1C}. However, these data often lack the  labels to connect their rich spectral information to specific evolutionary stages. The scale and complexity of these data sets make traditional analytical approaches increasingly ineffective, as manual classification becomes both time-consuming and prone to subjectivity. 

In recent years, machine learning (ML) algorithms have been increasingly applied to the analysis of complex observational data sets \citep[e.~g.][]{sen2022astronomical, 2011A&A...525A.157H, 2022A&A...661A...6S}. Unlike traditional statistical methods, ML techniques are capable of uncovering hidden dependencies and complex patterns that are difficult to detect using only statistical approach and several selected features. The advantage of ML is that it can reveal  multidimensional combinations of features that are not obvious when analysed by several parameters, which is especially important for complex data.

Current studies highlight the growing role of ML in astronomy, with applications ranging from galaxy classification to transient detection (\cite{2017ApJS..230...20A, 2023A&A...674A..13E, 2021PhRvL.127x1103D}). Machine learning techniques are increasingly used to enhance the precision of astrophysical parameter estimation. For instance, \cite{Meshcheryakov2023} applied ML within the SRGz system to identify and classify optical counterparts of SRG/eROSITA X-ray sources, yielding a detailed catalogue of source types and photometric redshifts and illustrating the power of ML to extract meaningful patterns from large multi-wavelength data sets. Furthermore, ML approaches have proven efficiency in the study of star formation. Star formation follows a continuous hierarchy ---  from diffuse molecular clouds to dense clumps, then to individual protostellar cores and young stellar objects, and finally to ionised \HII{} regions and surrounding photon-dominated regions (PDRs) shaped by stellar feedback. Characterising clumps along this sequence allows us to map physical and chemical transitions that govern the transformation from diffuse molecular gas to ionised star-forming regions. Recent study by \cite{2025A&C....5300985M}, have demonstrated the potential of ML to identify the evolutionary stages of Hi-GAL sources based on photometric and physical parameters, showing that ML can reproduce and even refine traditional, human-defined classification schemes. \cite{Angarita2025}  demonstrated the successful application of ML algorithms to millimetre spectra of massive young stellar objects (YSOs), identifying common spectral signatures linked to physical conditions in star-forming regions. By combining Gaia DR2 with WISE and Planck data and applying machine learning classification, \cite{2019MNRAS.487.2522M} constructed an all-sky probabilistic catalogue of YSOs and identified more than one million YSOs candidates. Together, these results underscore the growing effectiveness of ML in deriving physically interpretable results from extensive and complex observational data.

While ML has proven highly successful across many astrophysical domains, its application to the study of molecular clumps and early star formation is still developing.  Traditional classification schemes often rely on IR emission to distinguish between different evolutionary stages of molecular clumps \citep{2005ApJ...634L..57S, 2013A&A...556A..16G}. These approaches are highly valuable for interpreting the evolutionary stages of molecular clumps and have traditionally relied on expert interpretation of infrared emission. However,  a considerable fraction of clumps in surveys remain unclassified due to ambiguous, distance uncertainties or heavily obscured IR signatures. 

In contrast to IR emission, molecular line observations contain direct information about the physical conditions (e.g., density, temperature) and chemical composition of the gas, offering a more detailed view of clump properties. Despite this potential, molecular line data have rarely been exploited for evolutionary classification. This motivates the use of data-driven methods capable of identifying patterns and correlations beyond those accessible through manual classification.

In this work, we investigate the capability of ML methods for the automated identification and classification of molecular clumps at different evolutionary stages. This article aims to test whether ML approach can serve as an complementary way to explore evolutionary stages of clumps directly from molecular line data, specifically using the integrated intensities. For unsupervised analysis, we employ clustering algorithms to identify statistically distinct groupings of molecular clumps, which can then be examined for possible correspondence with different evolutionary stages. For supervised analysis  we used human-assigned evolutionary classifications provided by the MALT90 catalogue to extend classifications for sources previously labelled as uncertain.

The paper is organised as follows: in Section~\ref{sec:data}, we describe the data sets used in this study including the MALT90 molecular line survey and the complementary \textit{Spitzer} IR data. In Section~\ref{sec:methods} described the preprocessing steps and applied machine learning methods. In Section~\ref{sec:results}, we present the results of the unsupervised clustering and supervised classification analyses, discussing the physical interpretation of the identified clusters and the classification performance. Section~\ref{sec:discussion} provides a discussion of the results in the context of star formation and molecular evolution. Finally, Section~\ref{sec:conclusions} summarises our main findings.

\section{Data} \label{sec:data}
\subsection{MALT90 molecular line data}
We apply ML methods to the catalogue, based on the results of the MALT90 survey \citep[The Millimetre Astronomy Legacy Team 90~GHz, see][]{2013PASA...30...57J}, a large-scale millimetre-wavelength project that mapped molecular line emission at 90~GHz around  dense clumps identified by the ATLASGAL survey \citep[The APEX Telescope Large Area Survey of the Galaxy at 870~$\mu$m, see][]{2009A&A...504..415S}.  The MALT90 survey produced  $3^{\prime} \times 3^{\prime}$ maps of clumps, with a pixel size of $9^{\prime\prime}$, an angular resolution of  $38^{\prime\prime}$, and a spectral resolution of  0.11~km~s$^{-1}$. Spectra were extracted for each clump towards dust continuum peaks, then Gaussian-profile fitting was used to derive properties of emission lines, which were included to the catalogue. All the clumps are distributed along the Galactic plane ($-1^{\circ}<b<1^{\circ}$) spanning over a Galactic longitude in range of $0^{\circ}<l<20^{\circ}$ and also includes objects located in the Galaxy’s fourth quadrant ($284^{\circ}<l<360^{\circ}$). Kinematic distances and Galactocentric radii for MALT90 sources were determined by \cite{2017AJ....154..140W}, who found that the catalogue spans clumps from the Galactic Centre to Galactocentric radii of up to 15~kpc (see their Figure~20 and Figure~25).  

The resulting catalogue \citep[see][]{Rathborne2016} includes 269 features for 3556 molecular clumps. Each source has IR-based category defined using \textit{Spitzer} images spanning 3-24$\mu$m: Q for IR-quiescent clumps (IR dark), A for protostellar (either through extended 4.5~$\mu$m emission that indicates shocked gas, or through a compact 24~$\mu$m point source tracing the warm dust around the embedded star), C for compact \HII{} regions (compact, IR bright), H for extended \HII{} regions (extended IR bright), P for photo-dissociation regions (PDR; characterised by strong 8~$\mu$m emission from polycyclic aromatic hydrocarbons), and U for molecular clumps whose evolutionary stage is uncertain (IR signatures are obscured by foreground emission or are ambiguous; hereafter referred to as uncertain objects or uncertain clumps). This classification scheme is human-assigned and it is based solely on  \textit{Spitzer} IR fluxes. Molecular line emission data were obtained from the MALT90 survey by downloading the publicly available catalogue in CSV format through the VizieR database\footnote{\url{https://cdsarc.cds.unistra.fr/viz-bin/cat/J/other/PASA/33.30}}. The 870~$\mu$m flux density at the dust continuum peak, also included in the MALT90 catalogue, was extracted from the ATLASGAL survey \citep{2009A&A...504..415S}. 

After the analysis based only on the integrated intensities of the molecular lines around 90~GHz, we complemented these data with infrared photometry, that  was performed using the \textit{Spitzer  Heritage Archive}\footnote{\url{https://irsa.ipac.caltech.edu/applications/Spitzer/SHA/?__action=layout.showDropDown&}} to enrich the molecular data set with mid-infrared emission properties. For further analysis, we also incorporated 
data on the presence of Class~II~\ce{CH3OH} masers located within 5$^{\prime\prime}$ of the sources, obtained from the maser database \citep[see][]{2019AJ....158..233L} as well as dust temperature estimates derived by \citet{2015ApJ...815..130G}. 

We performed data analysis using the {\tt Scikit-learn} machine learning library for {\tt Python}~\citep{pedregosa2011scikit}.

\subsection{Photometry of MALT90 objects using \textit{Spitzer Heritage Archive}}

As mentioned in \cite{Rathborne2016}, infrared photometry data for MALT90 objects at wavelengths of 3–8~$\mu$m and 24~$\mu$m were taken from the GLIMPSE and MIPSGAL surveys~\citep[see][]{2003Benjamin, carey2009mipsgal}. However, despite the fact that the photometry data are available in the VizieR database~\footnote{\url{https://vizier.cds.unistra.fr/viz-bin/VizieR?-source=II/293}}, they cover only objects at the galactic longitude $|l|= 10^{\circ} -65^{\circ}$, which covers less than 2/3 of the clumps represented in MALT90 catalogue. At the same time, \cite{carey2009mipsgal} obtained mosaics for the galactic plane, but photometry at 24~$\mu$m was not carried out for our sources of interest. 

Therefore, photometric measurements at wavelengths of 3.6, 4.5 and 8~$\mu$m were performed independently using the archival data of the \textit{Spitzer  Space Telescope} IRAC (Infrared Array Camera) in channels 1, 2 and 4, respectively. The processing technique was based on the recommendations given in the IRAC Instrument Handbook~\footnote{\url{https://irsa.ipac.caltech.edu/data/SPITZER/docs/irac/iracinstrumenthandbook/IRAC_Instrument_Handbook.pdf}}. For each object from the MALT90 catalogue, aperture photometry of the mosaic images was carried out with subsequent correction of the background emission. The measurements were performed using a circular aperture with a radius of 2.4$^{\prime\prime}$. The background emission was estimated in a region with an inner radius of 2.4$^{\prime\prime}$ and an outer radius of 7.2$^{\prime\prime}$. To correctly account for flux losses outside the aperture, an aperture correction corresponding to the IRAC channel was applied. The background emission was estimated using the MMM (Mean-Mode-Mean) algorithm, {\tt MMMBackground} class from {\tt photutils} library \citep{larry_bradley_2025_14889440}, which provides an estimate of the sky background emission even in the presence of weak sources in the ring region. The measured flux in the aperture was corrected to subtract the background emission. 

Photometry at a wavelength of 24~$\mu$m  was performed in another way due to the larger beam size and the specific sensitivity of the MIPS (Multiband Imaging Photometer for Spitzer) camera. As for the IRAC instrument, there is a data processing manual for MIPS~\footnote{\url{https://irsa.ipac.caltech.edu/data/SPITZER/docs/mips/mipsinstrumenthandbook/MIPS_Instrument_Handbook.pdf}}, which was taken as a basis for performing the photometry. For each object, aperture photometry was performed with an aperture radius of 7$^{\prime\prime}$, and the background radiation was estimated in an annular region with an inner radius of 7$^{\prime\prime}$ and an outer radius of 20$^{\prime\prime}$. The sequence of data processing and flux calculation did not differ from that used for the IRAC channels. We managed to perform photometry for around 90\% of sources, as for some cases mosaic images were not present.  It is worth to mention, that by following the photometric procedures described in the handbooks, the obtained accuracy of the photometry results is reported to be 10--20\%. 

 After performing the photometry, we summarised the data set completeness in 
 Table~\ref{tab:nonzero_counts}. Photometric measurements could not be performed for all the sources due to the absence of mosaic images for some of them or because certain images were found to be corrupted.

\section{Methods} \label{sec:methods}
\subsection{Workflow of the machine learning analysis}\label{sec:ml_wf}

Clustering was performed in two approaches in order to explore the parameter space and identify natural groups among the sources. In the first approach, we restricted the parameter set to those detected in at least 40\% of sources, ending up with five molecular lines belonging to molecules \ce{HCO^+}, HNC, \ce{N2H^+}, HCN, and \ce{C2H} together with the 870~$\mu$m flux (see Table~\ref{tab:nonzero_counts}). In the second approach, we consequently added less common molecular emission, extending the analysis to the full set of MALT90 line intensities, together with the \textit{Spitzer} infrared fluxes derived in this work and additional features.

Furthermore, we explored supervised learning approaches to further investigate the evolutionary classification of the sources. In this case, the human-assigned classification from the MALT90 catalogue were used as training labels, assuming these labels to be a priori correct, allowing us to perform classification with supervised algorithms. The primary goal was to evaluate whether sources with uncertain or ambiguous evolutionary categories could be more clearly defined using the available observational data. By combining molecular line intensities from MALT90 and \textit{Spitzer} IR fluxes as features, we aimed to determine how effectively the algorithms could reproduce the established classifications.

\begin{figure*}
    \centering
    \includegraphics[width=2\columnwidth]{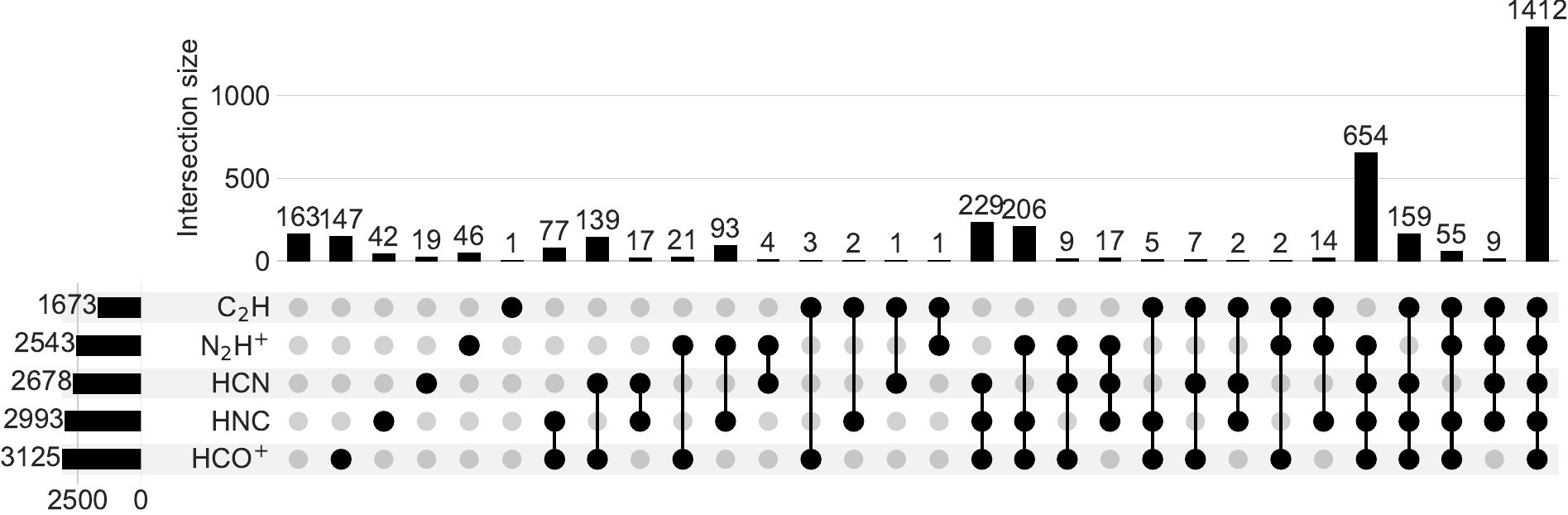}
    \caption{Detection patterns of five most common molecular lines across MALT90 catalogue. Each row represents a molecule, and each column represents a specific combination of detected molecules. Grey dots indicate non-detections, while black dots indicate detected lines. Black dots within a column are connected to show which molecules are detected together. The horizontal bars on the left indicate the number of sources in which each molecule is detected, while the vertical bars on top represent the number of sources with each specific combination of molecules ("intersection size").}
    \label{fig:upset}
\end{figure*}

\begin{table}
\caption{Summary of feature completeness for the data sets used in this work: molecular line intensities from MALT90 catalogue, \textit{Spitzer} IR fluxes after photometry was performed, and Class~II CH$_{3}$OH maser detections.}
\label{tab:nonzero_counts}
\begin{tabular}{lrr}
\hline
Feature & Non-zero count & Percentage (\%) \\
\hline
870 Flux peak & 3556 & 100.00 \\
8.0~$\mu$m                     &   3542  &   99.61  \\     
3.6~$\mu$m                     &   3535  &   99.41  \\     
4.5~$\mu$m                     &   3209  &   90.24  \\     
24~$\mu$m                      &   3208  &   90.21  \\     
HCO$^{+}$                   &   3125  &   87.88  \\     
HNC                         &   2993  &   84.17  \\     
HCN                         &   2678  &   75.31  \\     
N$_{2}$H$^{+}$              &   2543  &   71.51  \\     
C$_{2}$H                    &   1673  &   47.05  \\     
H$^{13}$CO$^{+}$            &   915   &   25.73  \\     
HC$_{3}$N                   &   798   &   22.44  \\     
HN$^{13}$C                  &   547   &   15.38  \\     
Maser CH$_{3}$OH            &   505   &   14.2   \\     
SiO                         &   341   &   9.59   \\     
HNCO            &   296   &   8.32   \\     
$^{13}$CS                   &   171   &   4.81   \\     
CH$_{3}$CN                  &   103   &   2.9    \\     
H41$_{a}$                   &   22    &   0.62   \\     
HC$^{13}$CCN                &   10    &   0.28   \\     
$^{13}$C$^{34}$S            &   9     &   0.25   \\ 
\hline
\end{tabular}
\end{table}

\subsection{Unsupervised learning techniques}
\label{sec:unsupervised}
\subsubsection{Dimensionality reduction techniques}

Clustering is a powerful technique to uncover patterns in high-dimensional data, allowing us to identify meaningful subgroups without prior labelling. However, when dealing with a large number of features, the presence of correlated, irrelevant or redundant information can distort the clustering structure, making interpretation more challenging. Additionally, not all features contribute equally to the formation of distinct subgroups. To address these challenges, dimensionality reduction techniques are often applied before clustering. 

\paragraph{Principal component analysis (PCA)}

PCA \citep{pearson1901, jolliffe2002} is one of dimensionality reduction techniques. It creates linear combination of features, transforming the data into a new set of variables, called principal components, which are ranked based on how much variance they capture. These components are linear combinations of the original features, with the first principal component (component~1) capturing the most variance, the second (component~2) capturing the next most, and so on. By reducing the data set to a smaller number of principal components, PCA helps remove noise and redundancy, making it easier to identify meaningful clusters. Mathematically this corresponds to a diagonalization of the covariance matrix, a rotation of the coordinate system in which the covariance matrix becomes diagonal. The eigenvectors define the directions of the principal components, while the eigenvalues represent their variances. By examining the rank of the covariance matrix and retaining only components with significant eigenvalues, PCA effectively reduces the dimensionality of the data set. This process can improve the performance of clustering algorithms by highlighting the most important features and reducing computational complexity. However, PCA has its limitations. This technique assumes that the data can be represented as a linear combination of components. If the data contains non-linear relationships, PCA may fail to capture them, which can lead to the loss of important information during clustering. In this study, PCA consistently produced a higher number of outliers and yielded lower silhouette scores. Therefore, we adopted an alternative dimensionality reduction technique to achieve a more reliable clustering representation.

\paragraph{t-distributed stochastic neighbour embedding~(t-SNE)}

Another of the dimensionality reduction techniques is t-SNE \citep{Maaten2008}. It is a nonlinear method designed to preserve local relationships between data points. t-SNE works by converting pairwise distances between points into conditional probabilities that represent similarities. It then tries to map the data into a lower-dimensional space, where similar points remain close together and dissimilar points are pushed farther apart. This approach often reveals underlying structures in the data.

However, a key limitation of t-SNE is that the resulting axes in the low-dimensional space do not correspond to any specific features or combinations of features from the original data set. While the distance between points tells us how similar they are, the direction, size, or exact placement of the whole plot doesn’t carry any interpretable meaning.

\subsubsection{Clustering algorithm HDBSCAN}

Unlike traditional clustering techniques, HDBSCAN does not require a predefined number of clusters. Instead, it dynamically identifies groupings based on local density variations. The algorithm also assigns some points as "noise", preventing artificial clusters from forming out of outliers. To perform clustering with this method we used {\tt HDBSCAN} library for {\tt Python}~\citep{mcinnes2017hdbscan}.

To prepare the data for clustering, we applied QuantileTransformer to standardize the features. Next, we employed the t-Distributed Stochastic Neighbor Embedding (t-SNE) technique for dimensionality reduction, which transformed the high-dimensional data into a two-dimensional space. It is important to note, that the resulting two-dimensional space is not derived from the properties of the data set, but rather is a predefined dimensionality, specified during the t-SNE transformation. This transformation is particularly useful for visualizing complex data structures.

We optimised the clustering parameters by systematically varying several settings. Specifically, we tested different data scaling methods, including StandardScaler, RobustScaler, MinMaxScaler, PowerTransformer, MaxAbsScaler, Normalizer and QuantileTransformer of the scikit-learn library \citep{pedregosa2011scikit}.  For the t-SNE dimensionality reduction, we varied the perplexity parameter from 30 to 100 with a step of 10.  In the HDBSCAN clustering algorithm, we explored minimum cluster size values from 30 to 150, and the minimum sample parameter was varied from half the cluster size upward in steps of 10, stopping before the next increment would exceed the cluster size itself. Another variable parameter was the number of components. Since t-SNE can generate a maximum of three components, we tested the two possible configurations, using either 2 or 3 components. Each unique combination of parameters was evaluated using the silhouette score, a standard metric for assessing clustering quality \citep{Rousseeuw1987}. The configuration that achieved the highest silhouette score was selected as the optimal setup, ensuring that the resulting clusters were well separated. We found that for our data QuantileTransformer scaler and t-SNE with 2 components systematically showed the better result of silhouette scores compared to other scalers or using 3-component case. Unless otherwise specified, the clustering results presented below assume the use of the QuantileTransformer scaler combined with 2 components of t-SNE.

\subsection{Supervised learning techniques}
\label{sec:supervised}

\subsubsection{Random Forest classifier}

To understand which features were driving the observed clustering, we trained a Random Forest (RF) classifier,  using the cluster assignments as target labels. This supervised learning approach, applied to the features space, helped us to determine the relative importance of each feature in forming the clusters.

Random Forest is a machine learning method used for classification and regression tasks~\citep{breiman2001random}. It operates by constructing a large number of decision trees during training and making predictions based on the majority vote (in classification) or average output (in regression) of the individual trees. Each tree is built using a random subset of the training data and a random subset of features, which helps to reduce overfitting and improve generalisation.

The method is particularly useful when dealing with complex, high-dimensional data sets, as it can capture nonlinear relationships and interactions between features. Moreover, Random Forest provides estimates of feature importance, which indicate how much each variable contributes to the model’s decision-making process. These values can provide information on what features play key roles in distinguishing between different classes.

In our analysis we used a Random Forest classifier for two main purposes. First, to identify the most important features contributing to the separation of clumps. Second, we used the classifier to predict the likely category of uncertain objects. To do this, we excluded the uncertain objects from the training set, trained the model on the pre-classified clumps, and then applied it to estimate the probability of each uncertain object belonging to a uncertain category.

\subsubsection{Gradient Boosting classifier}

Gradient Boosting is another supervised machine learning method used for both classification and regression problems~\citep{friedman2001greedy}. Unlike Random Forest, which builds many independent trees, Gradient Boosting builds trees sequentially, where each new tree tries to correct the errors made by the previous ones. This is done by fitting each tree to the gradient of the loss function, focusing on the residuals or mistakes left by earlier trees.

The model gradually improves by adding these trees one at a time, combining their outputs to create a strong overall predictor. This sequential learning process allows Gradient Boosting to capture complex patterns and interactions in the data. Gradient Boosting also offers flexibility in terms of loss functions and regularisation techniques, helping to reduce overfitting and improve generalisation.

We applied a Gradient Boosting classifier~\citep{pedregosa2011scikit} in a similar manner as RF. We used the trained model to classify uncertain objects by excluding them from the training set, fitting the classifier to the known clumps, and then estimating the probability of each uncertain object belonging to a given class.

\section{Results}\label{sec:results}

\subsection{Clustering based on emission of the 5 molecular lines}

We note that not all molecular lines were detected toward every object. For lines with fluxes below the detection threshold, we included them in the analysis with an intensity of zero. Since the observations were blind within a frequency range around 90~GHz, a non-detection still provides meaningful information, meaning that the line is absent or very weak in that object. Assigning zero therefore allows the clustering algorithm to incorporate this information.

\begin{figure}
    \centering
    \includegraphics[width=1\linewidth]{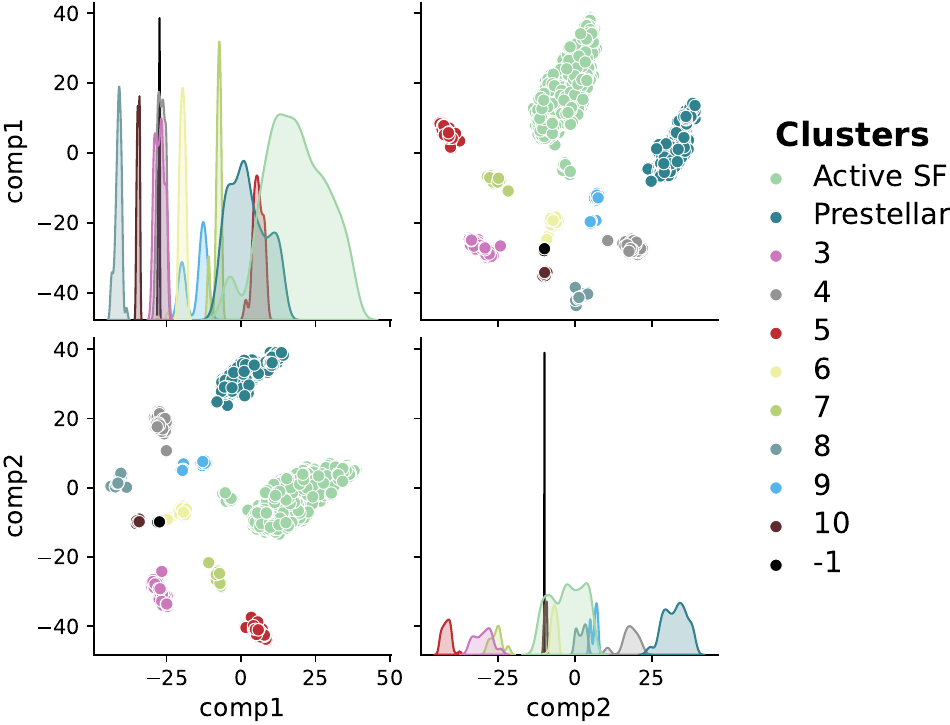}
    \caption{Result of HDBSCAN clustering based on integrated intensity values of five molecules: \ce{HCO^+}, HNC, \ce{N2H^+}, HCN and \ce{C2H}. Clusters are shown in different colours, with stable clusters labelled according to the most abundant category within each cluster. The t-SNE axes (comp1 and comp2) represent t-SNE result  dimensions. The legend shows clusters in the descending order and the cluster labelled as -1 represent outliers. The main diagonal shows the distributions of the residual dimensions for the clusters, represented as kernel density estimate (KDEs). Peaks roughly indicate modes, and the spread provides a visual sense of variance.}
    \label{fig:clusters_1}
\end{figure}

Using HDBSCAN, we successfully identified eleven clusters, displayed in Figure~\ref{fig:clusters_1}, each representing a distinct grouping within the data set based on the integrated intensities of C$_2$H, \ce{N2H+}, HCN, HNC, and \ce{HCO+}. Each dot corresponds to an object from the MALT90 catalogue. The highest silhouette score of 0.68 (maximum possible value is unity) was achieved with a minimum cluster size of 75 and a minimum sample size of 45. The most populated clusters are 1 and 2, indicating a higher concentration of data points within these groups compared to the others. As shown in Figure~\ref{fig:clusters_1}, while there is some overlap between the two clusters in the first component (comp1), the second component (comp2) remains distinct. This distinction allows for a clear separation of the clusters.

\begin{figure}
    \centering
    \includegraphics[width=0.8\linewidth]{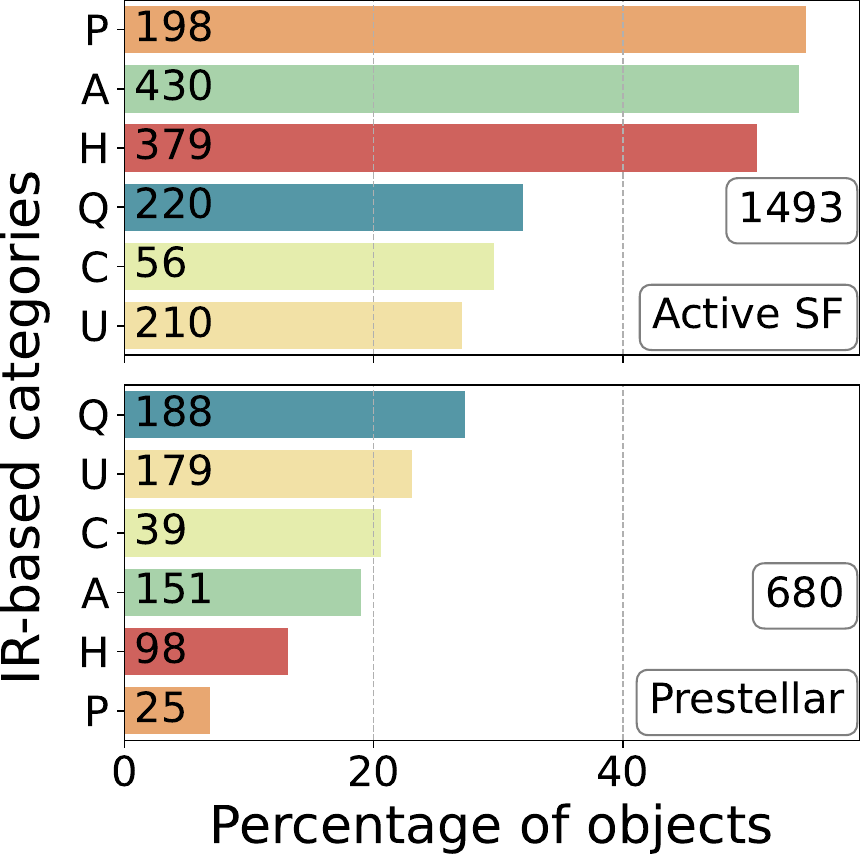}
    \caption{The distribution of object types within the two most stable and populated clusters identified with integrated intensities values of five molecules: \ce{HCO^+}, HNC, \ce{N2H^+}, HCN and \ce{C2H}.  Total number of objects in the each cluster is shown in the bottom right corners of the panels. Each bar represents a certain object type from MALT90 catalogue: Q for quiescent, A for protostellar, C for compact regions, H for extended regions, P for photo-dissociation regions and U for uncertain classifications. The percentage indicate the fraction of certain object type within a cluster, while he actual number of objects is shown along the x-axis.}
    \label{fig:clusters_1_2}
\end{figure}

Cluster 1 predominantly contains protostellar objects (A), PDRs (P), and extended \HII{} regions (H). We refer to this cluster, which consists predominantly of such objects, as Active star formation ( Active SF) cluster. Cluster 2 consists mainly of IR-quiescent sources (Q) and molecular clouds with uncertain classification (U). Therefore, we refer to this cluster as Prestellar in the following analysis. The composition of these two clusters is given in Figure~\ref{fig:clusters_1_2}.

\begin{figure}
    \centering
    \includegraphics[width=0.8\linewidth]{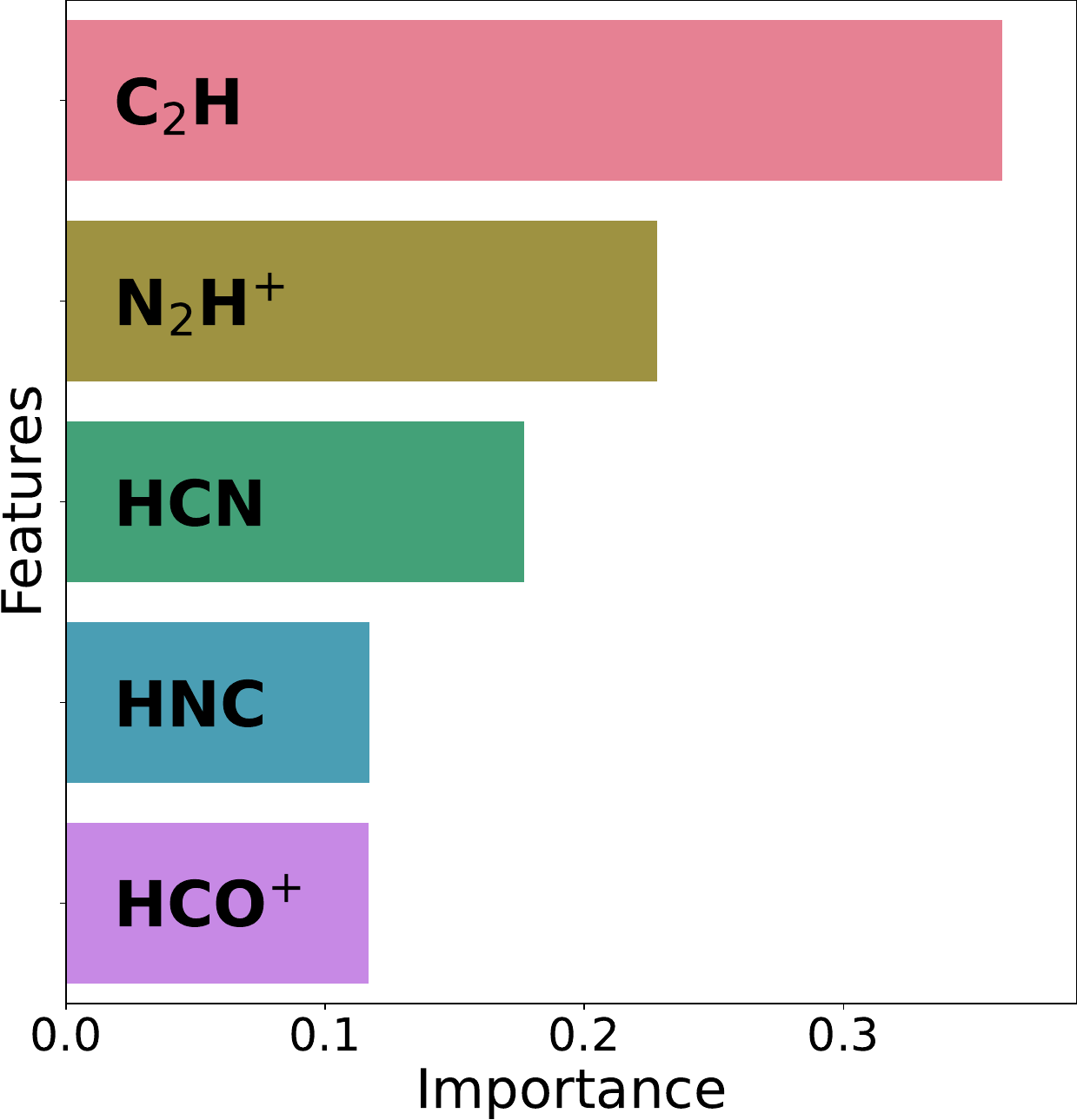}
    \caption{Random Forest-derived feature importance for the classification model based on integrated intensities of five molecular lines.}
    \label{fig:clusters_1_3}
\end{figure}

Since t-SNE is a nonlinear method, it is not possible to directly identify the contribution of individual features to the clustering result. To address this, we trained a Random Forest classifier on the data set, which allowed us to estimate feature importance based on how effectively each feature predicts cluster membership. The results, presented in Figure~\ref{fig:clusters_1_3}, show that the most important features are the integrated intensities of C$_2$H and \ce{N2H+}.

\subsection{Clustering based on emission of the 5 molecular lines, continuum at 870~$\mu$m and dust temperature}

Our next step was adding flux peak values at 870~$\mu$m as an extra feature to check how much it affects clustering. However, this feature does not significantly help to distinguish these object types and does not have any effect on the classification. As a result, the overall clustering remained the same, confirming the stability of the original result. 

While dust temperature was included as an additional feature in the clustering analyses, its contribution to the results was found to be minor. For this reason, we do not present separate results based on dust temperature, but instead discuss possible reasons for its limited impact in Section~\ref{sec:supervised}. We chose not to keep flux peak values at 870~$\mu$m  and dust temperature in our future clustering experiments in order to avoid potential contamination.

\subsection{Clustering based on \textit{Spitzer} fluxes with and without emission of the  molecular lines}

We also performed clustering based on fluxes of corresponded objects on \textit{Spitzer} data at 3.6, 4.5, 8.0, and 24~$\mu$m wavelengths and explored whether they provide separation of object categories. For objects lacking measurements in one or more \textit{Spitzer} bands, we used the median flux value of the corresponding  IR category of the clump.  However, the resulting silhouette scores did not exceed 0.3, implying that the data set only based on the fluxes measured by {\it Spitzer} does not provide a robust basis for classification.

The next step in our analysis involved combining the \textit{Spitzer}  IR fluxes with the integrated intensities of molecular lines from the MALT90 survey. We found that the inclusion of IR fluxes had minimal impact on the clustering results, indicating that their contribution to cluster separation was low.

\subsection{Clustering based on molecular emission, \textit{Spitzer} IR fluxes and methanol masers indicators} \label{sec:ii+IR}

Subsequently, we explored the effect of adding individual molecular features from Table~\ref{tab:nonzero_counts}. We found that the addition of H$^{13}$CO$^+$ led to the emergence of a distinct cluster characterised by \HII{} regions and PDR. 

\begin{figure}
    \centering
    \includegraphics[width=1\linewidth]{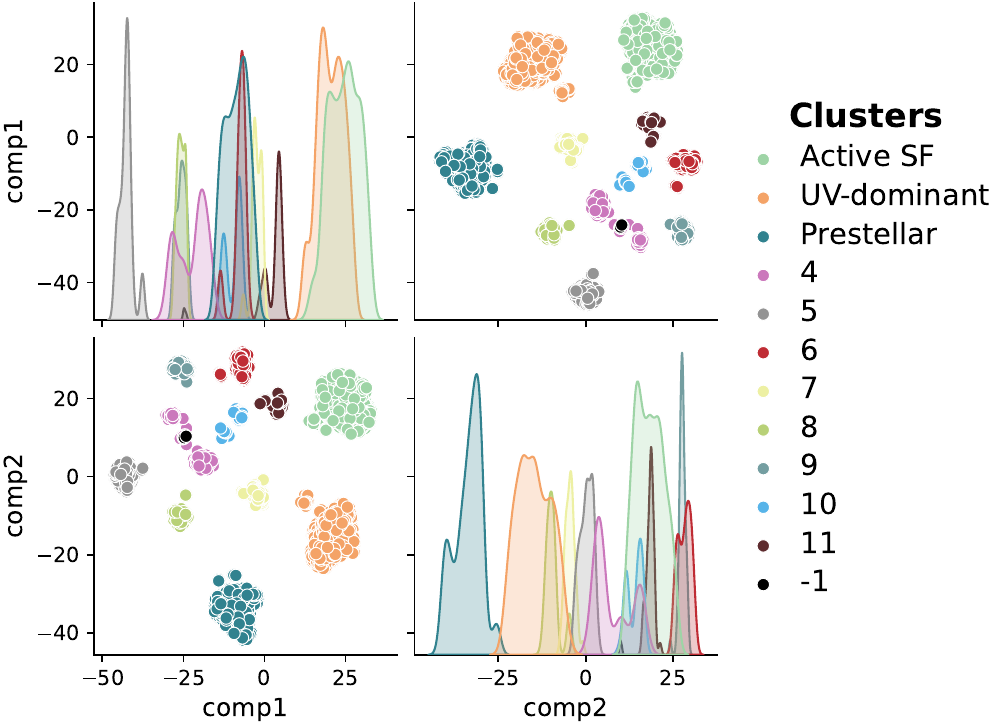}
    \caption{Result of HDBSCAN clustering 
    based on integrated intensity values of six molecules: \ce{HCO^+}, HNC, \ce{N2H^+}, HCN and \ce{C2H}, \ce{H^13CO^+} and {\it Spitzer} IR emission. Comp1 and comp2 do not correspond to those in Figure~\ref{fig:clusters_1}}
    \label{fig:clusters_IR}
\end{figure}

In Figure~\ref{fig:clusters_IR} we show clustering results obtained using integrated intensities of six molecular emission lines in combination with {\it Spitzer} IR fluxes. It shows three distinct clusters. A silhouette score of 0.71 was achieved using the QuantileTransformer for scaling, a t-SNE perplexity of 90, a minimum cluster size of 70, and a minimum sample size of 55. The legend shows clusters in the descending order and the cluster labelled as -1 represent outliers (247 objects).

\begin{figure}
    \centering
    \includegraphics[width=0.8\linewidth]{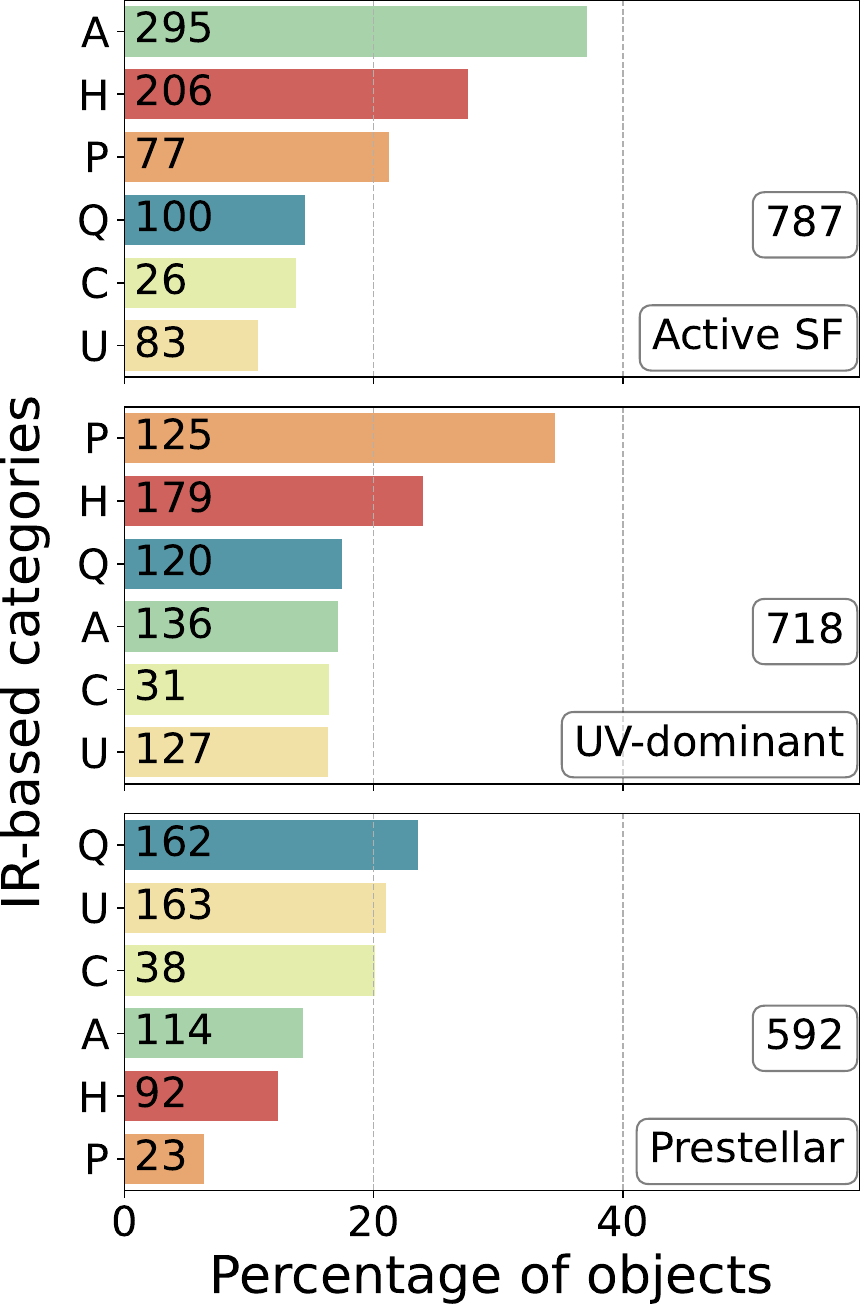}
    \caption{The distribution of object types within the three most stable and populated clusters identified using HDBSCAN clustering with t-SNE dimensionality reduction on integrated intensities values of six molecules: \ce{HCO^+}, HNC, \ce{N2H^+}, HCN and \ce{C2H}, \ce{H^13CO^+}, and {\it Spitzer} IR emission. The meaning of all the number is the same as in Figure~\ref{fig:clusters_1_2}.}
    \label{fig:classification_IR}
\end{figure}

Detailed distribution of the IR-based categories within each cluster is displayed in Figure~\ref{fig:classification_IR}. The most populated cluster, named Active SF, contains 787 sources, of which 501 are protostellar objects and \HII{} regions. This composition indicates that the cluster mainly represents the intermediate to advanced stages of star formation. 

The UV-dominant cluster is less populated at $\approx 10\%$ and contains 718 objects. The cluster includes a significant fraction of more evolved sources: approximately 34.5\% of all PDRs and 24\% of all \HII{} regions from the MALT90 catalogue are associated with this cluster. This distribution indicates that the second cluster probably corresponds to the regions where the star formation process has progressed further, leading to more prominent photo-dissociation and ionisation features. 

The third cluster, referred to as Prestellar contains 592 clumps and it is dominated by sources without clear signatures of active star formation. It includes 23.5\% of all IR-classified quiescent (Q) clumps and 21\% of the objects with uncertain evolutionary stage (U). The comparison between the clusters in this case and when using only five molecular lines are discussed in detail in the Section~\ref{sec:discussion}

\begin{figure}
    \centering
    \includegraphics[width=0.7\linewidth]{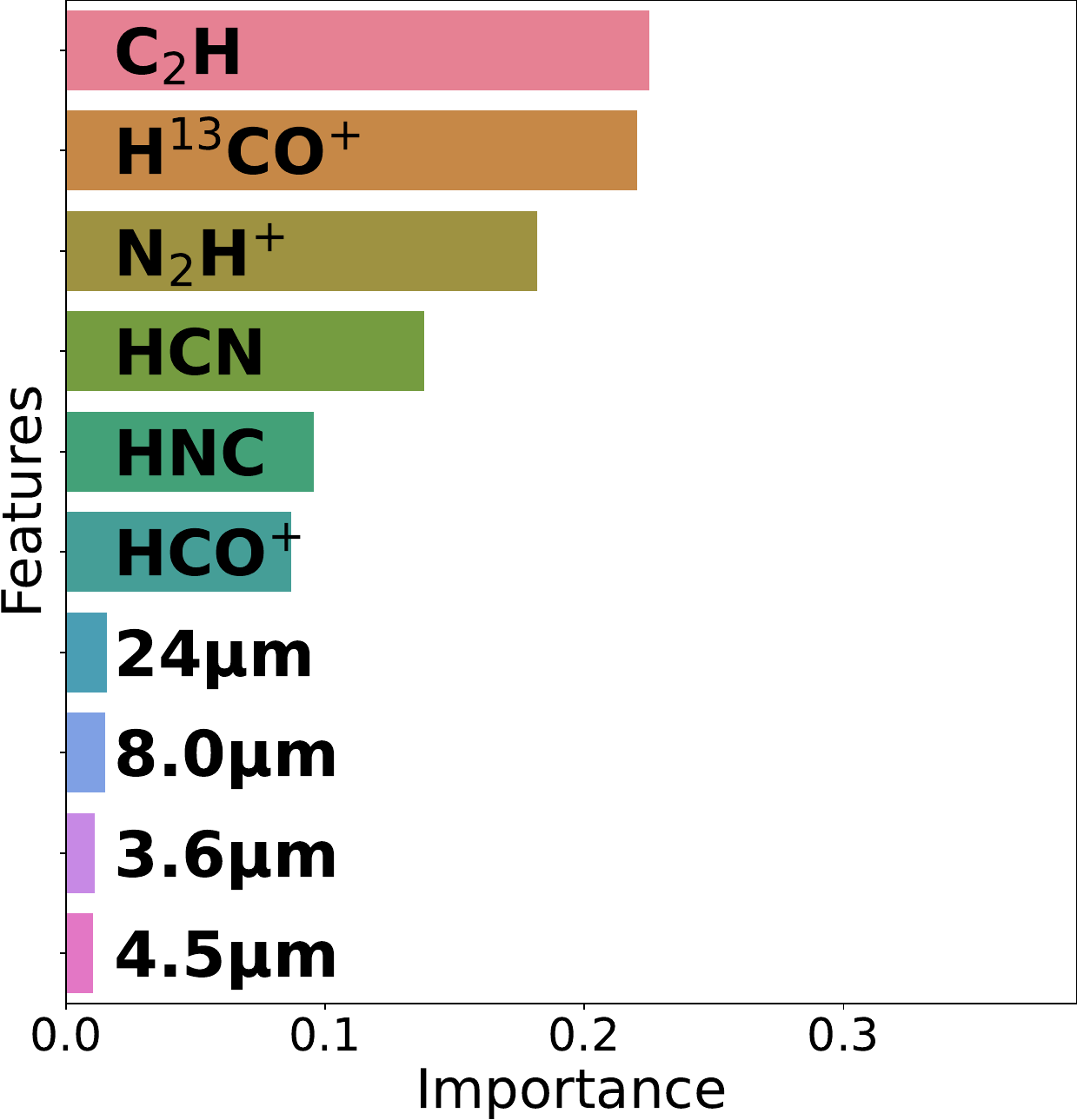}
    \caption{ Random Forest-derived feature importance for the classification model based on integrated intensities of molecular lines and {\it Spitzer} IR fluxes}
    \label{fig:importance_IR}
\end{figure}

The most important features for the obtained classification are the integrated intensities of C$_2$H, H$^{13}$CO$^+$ and \ce{N2H+} molecules, see Figure~\ref{fig:importance_IR}. IR fluxes from {\it Spitzer} were not the most important features which allowed separating the UV-dominant cluster from the data set. Instead, the H$^{13}$CO$^+$ line emission became one of the main features.

Incorporating other molecular tracers generally resulted in further fragmentation of existing clusters into smaller, more specific subgroups. For example, the inclusion of methanol maser indicators caused the cluster dominated by protostellar objects and \HII{} regions (categories A and H) to split into two distinct groups: one associated with maser emission and another without and this is a trivial result. 

Addition of other molecular emission lines did not lead to the emergence of new stable clusters; the clustering consistently produced groups dominated by \HII+Protostars, \HII+PDR, and Quiescent+Uncertain objects in varying proportions. 

\subsection{Effect of feature completeness on clustering results analysis}
 Although we refer to feature completeness, we note that the absence of detected emission in a given molecular transition reflects a true non-detection and therefore contributes information about the physical state of the medium. To assess the robustness of the clustering results with respect to variations in feature completeness, we performed an additional analysis on a restricted “high-information” subset of the catalogue. 

This subset consists of 2066 molecular clumps with detections in the four most frequently detected MALT90 molecular transitions (\ce{HCO+},HNC, HCN, \ce{N2H+}) and complete \textit{Spitzer} infrared flux measurements (at 3.6, 4.5, 8, and 24~$\mu$m). For these sources, a total of 20 features were used, comprising four mid-infrared fluxes and 16 velocity-integrated molecular line intensities. The clustering was carried out using the same methodology as applied to the full catalogue. 

The resulting cluster structure is qualitatively consistent with that obtained from the full sample with three dominant clusters emerging, that broadly correspond to UV-dominant, Active SF and Prestelllar evolutionary clusters. While the relative populations of the clusters differ due to the more restrictive selection, the overall clustering behaviour and physical interpretation remain unchanged. Feature-importance analysis indicates that the clustering in this subset is primarily driven by the molecular tracers, which together account for approximately 75\% of the total feature importance.

\subsection{Machine learning classification of uncertain clumps}

More than 20\% of the sources make the clumps with uncertain evolutionary stage in the MALT90 catalogue. As shown in Figure~\ref{fig:clusters_1_2} and Figure~\ref{fig:classification_IR}, a significant portion of the Prestellar cluster consists of Uncertain objects, making them one of the major components of that group. These reasons motivated us to explore whether it is possible to identify the evolutionary stage of Uncertain clumps using supervised machine learning techniques.

We started with preliminary data reduction, that was described in Sec.~\ref{sec:methods}. As in the clustering analysis we also applied QuantileTransformer scaler which had previously showed the best results on the data set we have. We used integrated intensities of \ce{HCO^+}, HNC, \ce{N2H^+}, HCN and \ce{C2H}, \ce{H^13CO^+} along with \textit{Spitzer}  IR emission at 3.6~$\mu$m, 4.5~$\mu$m, 8.0~$\mu$m and 24~$\mu$m. Fluxes were obtained for 90–99\% of objects from the catalogue and for sources where fluxes were missing, we used the median flux value corresponding to the same object type. The data were then split into two groups with different sizes: objects classified by \citet{Rathborne2016} and those for which evolutionary category remained uncertain. We trained a Random Forest classifier with 1000 trees using the subset of data with known categories to learn underlying patterns. To evaluate the model’s performance and ensure its reliability, we conducted 4-fold cross-validation. In this procedure, the data set was divided into four equally sized sets (folds) by splitting the objects (rows) while keeping all their features (columns) intact. During each iteration, three folds were used to train the model, while the remaining fold was used for validation. This process was repeated four times so that each fold served once as the validation set. The results from all folds were then averaged to obtain a more robust and unbiased estimate of the model's generalisation performance. Then we used the trained model to predict the most likely category for objects in the uncertain group, along with the probability of belonging to each category.

The average cross-validation accuracy with the Random Forest classifier appeared to be 0.61, which is moderate, therefore, we decided to implement a threshold of  50\% probability for belonging to the predicted category. If the probability of an uncertain object belonging to the predicted evolutionary category was lower than this threshold, we assumed that the object remains uncertain. With this approach, we identified 550 out of 774 objects in the uncertain category. The summarised result is shown in Table~\ref{tab:URF_GB}.

\begin{table}
\caption{Summary of predicted object category for uncertain clumps with Random Forest (RF) and Gradient Boosting (GB) classifiers. Statistics is shown for predictions with probability $> 0.5$.}
\label{tab:URF_GB}
\begin{tabular}{ccccccc}
\hline
\shortstack{Predicted\\category}   & \multicolumn{2}{c}
{\shortstack{Number\\ of objects}}   & \multicolumn{2}{c}{\shortstack{Percentage \\ of total(\%)}} & \multicolumn{2}{c}{\shortstack{Average \\ probability}}\\
  & RF& GB&     RF& GB&     RF& GB \\
\hline
Q & 457  & 476  &  83.1& 77.9    &   0.75& 0.77 \\
P &  49  &  73  &   8.9  &11.9    &  0.67 &0.72 \\
A &  39  &  57  &   7.1  & 9.3    &  0.60 &0.61 \\
H &   5  &   5  &   0.9  & 0.8   &   0.61 & 0.60\\
C &   --  &  -- &    --  &   -- &     -- &  --\\
\hline
\end{tabular}
\end{table}

Since cross-validation accuracy wasn't sufficient we decided to implement Gradient Boosting -- another method to predict category for uncertain objects from the data set. We used a Gradient Boosting classifier with 1000 estimators, a learning rate of 0.01, and a maximum depth of 3. To improve generalisation and reduce overfitting, we applied a subsampling rate of 0.8. Additionally, we enabled early stopping with a validation split of 10\%, allowing training to halt automatically if performance did not improve over 10 consecutive iterations. The model was trained on the subset of data with defined categories, which were previously assigned by \cite{Rathborne2016}. To evaluate the model's performance, we applied 4-fold cross-validation  consistent with the approach used for the Random Forest model.

\begin{figure}
    \centering
    \includegraphics[width=0.7\linewidth]{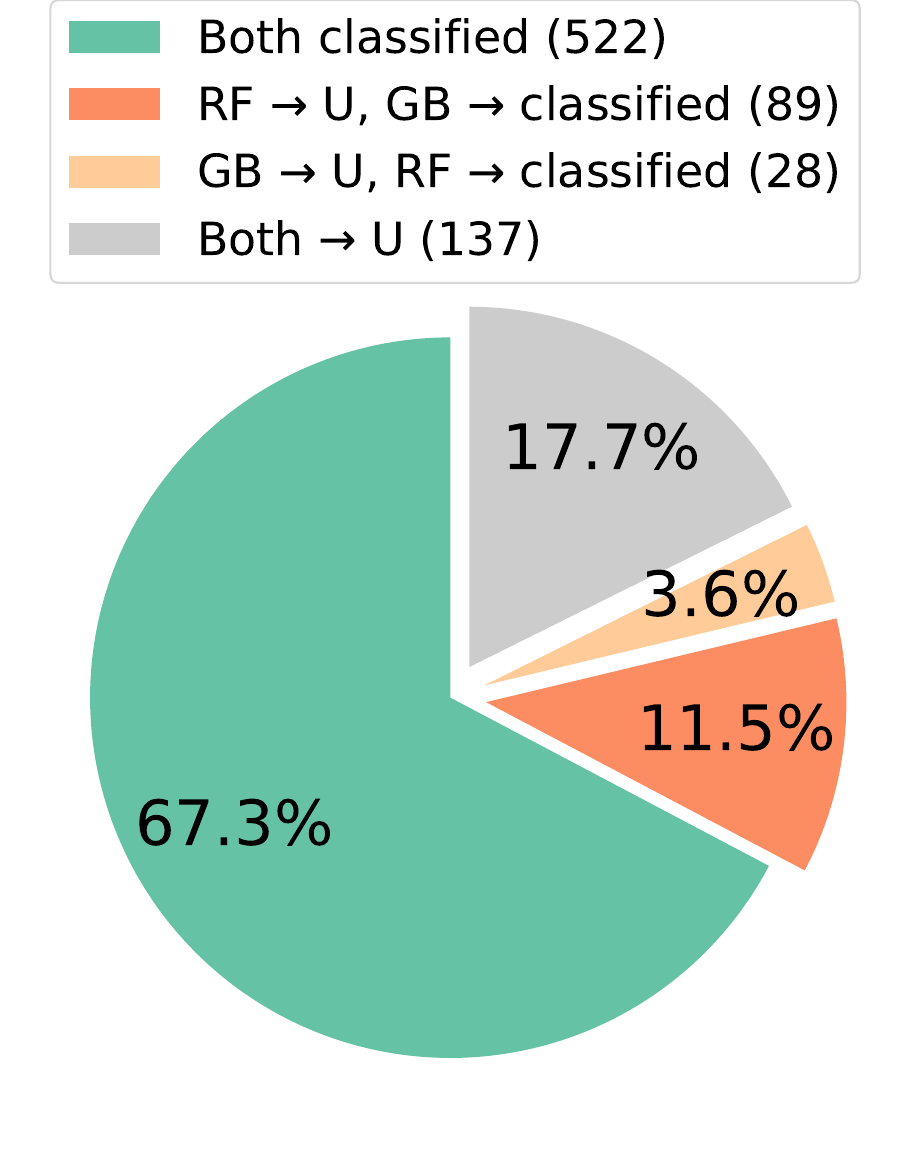}
    \caption{Comparison of classification outcomes for previously uncertain objects~(U) using Random Forest~(RF) and Gradient Boosting~(GB) classifiers. The chart highlights agreement, disagreement and cases where classification was not confident. Each segment shows the number and percentage of objects classified with a probability $>0.5$ by one or both methods.}
    \label{fig:comparison_rf_gb}
\end{figure}

The cross-validation accuracy for the Gradient Boosting method was comparable to that of the Random Forest, reaching 0.61. Therefore, we applied a 50\% probability threshold for category assignment and retained classifications only for clumps with consistent classifications from both methods to ensure a robust final predictions. As a result, 512 objects were assigned to the same category by both methods (27 Protostars + 4 \HII{} regions + 45 PDRs + 436 IR-quiescent). A summary of how all uncertain objects were classified by the two models is presented in Figure~\ref{fig:comparison_rf_gb} as a pie chart.

\begin{figure}
    \centering
    \includegraphics[width=1\linewidth]{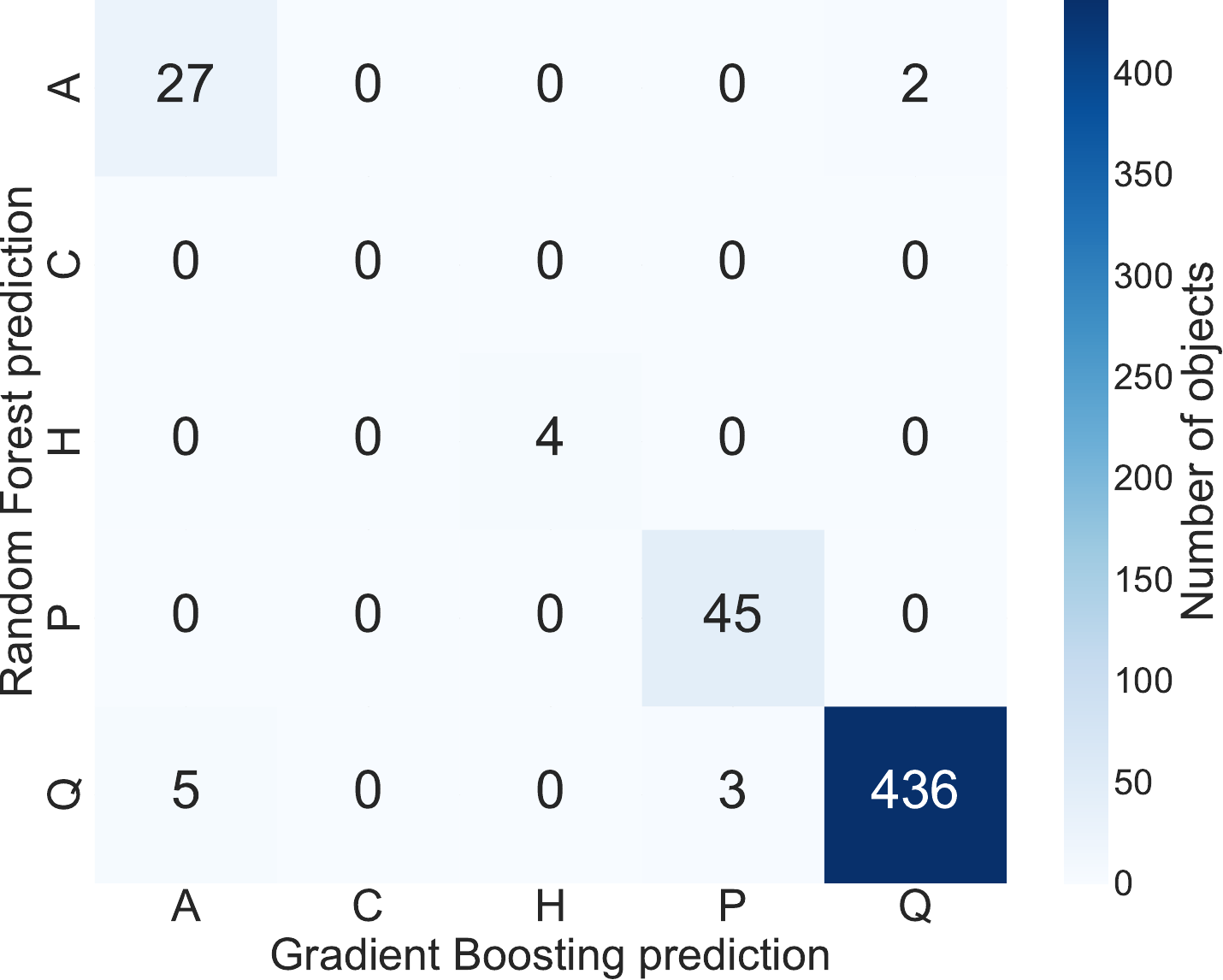}
    \caption{Classifier comparison  matrix illustrating agreements and disagreement between predictions made by the Random Forest and Gradient Boosting models. Each cell represents the number of objects for which the two classifiers assigned labels, with the colour intensity indicating the count.}
    \label{fig:confusion_matrix_rf_gb}
\end{figure}

To illustrate the differences in classification outcomes for the remaining uncertain cases, we present a classifier comparison matrix in Figure~\ref{fig:confusion_matrix_rf_gb}. Each cell indicates the number of objects for which the Random Forest (rows) and Gradient Boosting (columns) predicted a particular category. Areas of agreements and disagreements between the two classifiers are marked with both colour intensity and overlaid numbers representing the object count in each cell. The plot reveals that only 10 molecular clumps were assigned to different categories by the Gradient Boosting and Random Forest classifiers with the most disagreements occurring between the protostellar~(A) and IR-quiescent~(Q) categories.

Classification results for 522 classified clumps with probability greater than 0.5 in both methods are presented in Table~\ref{tab:classification_results}. We note that these clumps include 512 objects of the same category and 10 objects which GB and RF classified differently. Most of these clumps were consistently assigned to the same category by both the Random Forest and Gradient Boosting classifiers. The clumps that were classified differently are marked with an asterisk. 

\begin{figure}
    \centering
    \includegraphics[width=0.7\linewidth]{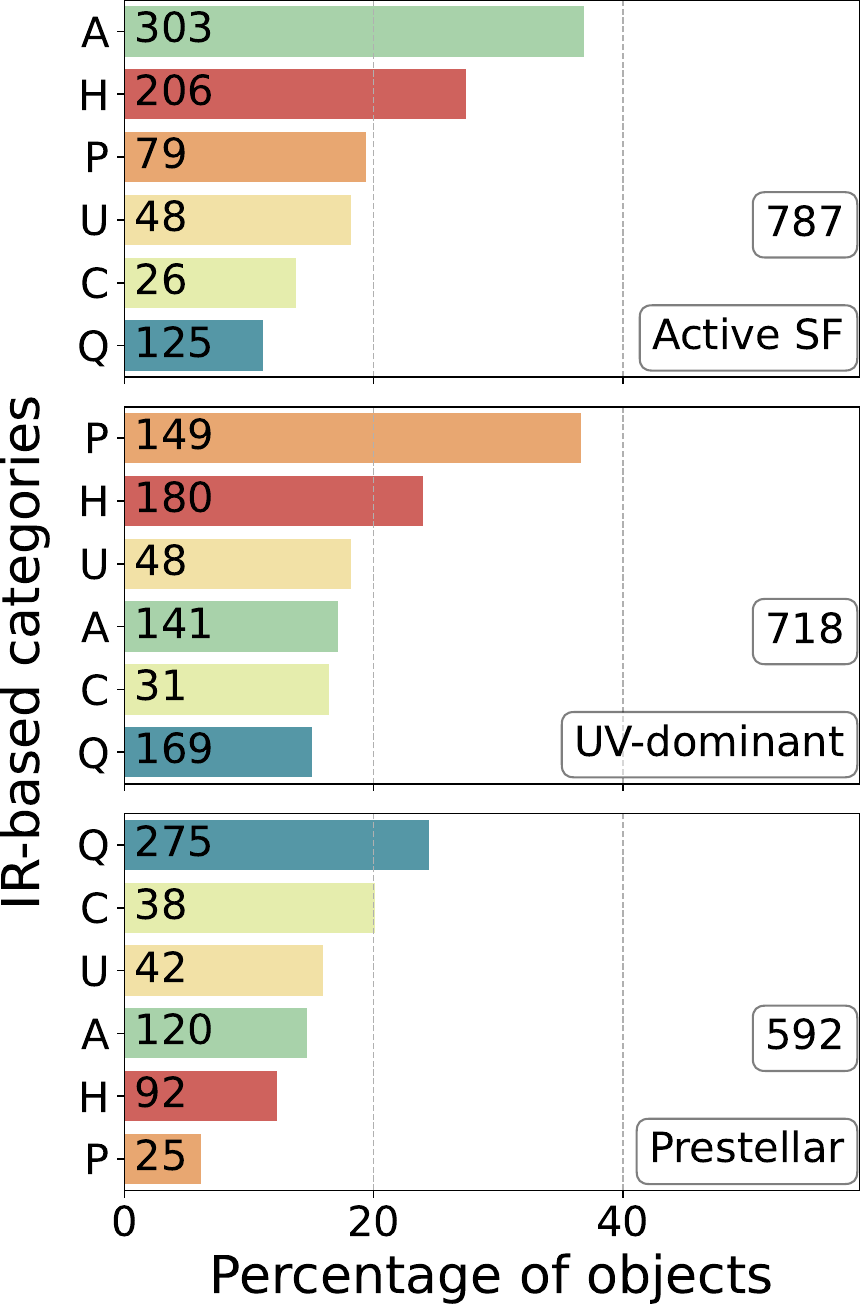}
    \caption{The distribution of object types within the three most stable and populated clusters after reassigning uncertain clumps (U) with the most probable category found from supervised learning. Clustering was performed using the same parameters as described in section~\ref{sec:ii+IR}. Compare this plot with Figure~\ref{fig:classification_IR}.}
    \label{fig:URF_GB_IR}
\end{figure}

Next, molecular clumps with the uncertain category were replaced with classifications from the RF and GB models. Clustering was repeated using the same parameters to assess the impact of the refined classifications. Refining the uncertain molecular clumps classification with ML clarified clustering results (see Figure~\ref{fig:URF_GB_IR}): clusters dominated by quiescent clumps became more distinct, while protostar-dominated clusters and \HII{}+PDRs remained stable. This demonstrates that ML-based classification enhances cluster homogeneity and highlights evolutionary stage patterns. Moreover, the refined classification is consistent with the structure revealed by the initial unsupervised clustering, indicating that the machine learning approach does not contradict, but rather reinforces previously identified clusters.

\section{Discussion}\label{sec:discussion}
In this study, we used data from the MALT90 survey and enhanced it with the \textit{Spitzer}  IR-fluxes measurements. We applied machine learning techniques with two main goals: (i) to examine whether algorithms, using only the integrated molecular line intensities and without prior knowledge of the sources, could identify the evolutionary stages of molecular clumps; and (ii) to evaluate whether models trained on molecular clumps with established IR-based categories could predict the evolutionary stages of clumps with uncertain classification.

\subsection{Unsupervised learning}
    Unsupervised learning applied to the molecular line data set revealed, that even when using only the integrated molecular intensities of five molecules, it is possible to distinguish two well-defined clusters corresponding to distinct evolutionary stages of molecular clumps (see Figure \ref{fig:clusters_1_2}). These include clumps without active signatures of star formation (Prestellar cluster) and clumps containing embedded protostars (Active SF cluster). However, this limited set of features does not allow identification of finer or more detailed group structures within the data set. 

The main features driving the clustering were the integrated intensities of C$_2$H and N$_2$H$^{+}$. These molecules trace environment with different physical conditions. \citet{2017A&A...599A..98P} found N$_2$H$^{+}$ the only reliable tracer of dense gas at 3~mm, that is resistant to depletion in early, pre-stellar environments. Its emission typically peaks in quiescent clumps prior to the onset of significant heating \citep{2012ApJ...756...60S, 2018ApJ...865..135Y}. In contrast, C$_2$H emission is weaker or absent in many cold clumps \citep{2019A&A...622A..32L} and becomes more prominent in outer or more UV-exposed regions \citep{2015ApJ...808..114J}. \citet{2017A&A...599A.100G} performed PCA analysis of integrated intensities at 3~mm and physical parameters of the Orion B cloud. They found positive contribution of the C$_2$H line intensity to the component estimating UV radiation. The contrasting behaviour of these tracers effectively captures the chemical and physical evolution of molecular clumps, allowing the unsupervised algorithm to distinguish between Prestellar and Active SF clusters. This behaviour finds reflection in Table~\ref{tab:ii_cluster_stat}, integrated intensities of  C$_2$H is under the detection threshold in the Prestellar  cluster, while it is present in Active SF cluster. Regarding to all other analysed lines, their intensities are higher up to 50-80\% in the Active SF cluster.

The 870~$\mu$m continuum emission did not noticeably influence the clustering results. This wavelength primarily traces the column density of cold dust, which varies smoothly across evolutionary stages and therefore provides limited contrast between prestellar regions and those with active star formation. Unlike molecular line intensities, which respond more sensitively to changes in temperature or chemistry, the 870~$\mu$m emission mainly reflects the total mass of the clump rather than its current physical and chemical state. Consequently, it offers little discriminative power for separating clusters.

\begin{table}
    \centering
\begin{tabular}{llrr}
\hline
Cluster & Feature & Mean & Median  \\
\hline
\multirow{5}{*}{Active SF} & HCO$^{+}$ & 7.26 & 4.54  \\
& HNC  & 5.14 & 3.15 \\
&  N$_{2}$H$^{+}$ & 6.19 & 4.10  \\
& HCN  & 8.36 & 4.72  \\
& C$_{2}$H  & 2.59 & 1.72  \\
\hline
\multirow{5}{*}{Prestellar}& HCO$^{+}$ & 5.73 & 2.83  \\
& HNC & 3.60 & 2.07  \\
&  N$_{2}$H$^{+}$ & 3.30 & 2.31  \\
& HCN & 7.27 & 2.63 \\
& C$_{2}$H & 0.00 & 0.00 \\
\hline
\end{tabular}
    \caption{Mean and median values of integrated molecular line intensities (K km s$^{-1}$) for molecular clumps associated with the  Active SF, and Prestellar clusters identified in this work.}
    \label{tab:ii_cluster_stat}
\end{table}
We also extended the analysis by incorporating additional integrated intensities that are less frequently detected in the MALT90 data set, together with \textit{Spitzer}  IR fluxes and  \ce{CH3OH} Class~II maser detection flags. The first notable difference emerged when the integrated intensity of H$^{13}$CO$^{+}$ was included, which produced a clearer separation of the cluster associated with \HII{} regions and PDRs. We therefore present the results for this configuration, since the inclusion of additional, less frequently detected molecular lines produced qualitatively similar outcomes. The analysis reveals three stable clusters: Prestellar, Active SF, and UV-dominant, dominated by IR-quiescent clumps, protostellar sources, and \HII{}+PDR regions, respectively. The only additional distinction appeared when CH$_3$OH maser flags were added. In that case, the Active SF cluster split into two subgroups: one containing sources with detected maser emission and another without, suggesting that adding CH$_3$OH maser detection flags helps the algorithm efficiently extract more active, high-mass star-forming regions within the protostellar population \citep{1987Natur.326...49B, 1991ApJ...380L..75M, 2018A&A...617A..80S}.

The clearer appearance of the UV-dominant cluster was primarily linked to the inclusion of the H$^{13}$CO$^{+}$ line. \textit{Spitzer}  IR fluxes  did not significantly affect the clustering pattern, whereas adding H$^{13}$CO$^{+}$ line allowed this group to emerge distinctly. This behaviour can be explained by the physical condition of the environment that affects the H$^{13}$CO$^{+}$ molecule. The molecule is destroyed by UV radiation in \HII{} regions and PDRs, and since H$^{13}$CO$^{+}$ forms under similar physical conditions as HCO$^{+}$ but involves the less abundant $^{13}$C isotope, its emission is generally weaker. As a result, H$^{13}$CO$^{+}$ emission is absent in both UV-dominated and IR-dark environments, but strongest in the intermediate, actively star-forming phase. Thus, even though H$^{13}$CO$^{+}$ emission itself is not strong within the UV-dominant cluster, its absence, combined with enhanced IR emission and the presence of UV-tracing molecules such as C$_2$H, provides a chemical contrast that helps the algorithm clearly separate this group.

As shown in Table~\ref{tab:cluster_stat} the Active SF cluster exhibits relatively high integrated intensities values. The infrared fluxes are also notably high, especially at 8.0 $\mu$m (mean 457.6, median 28.1 mJy) and 24 $\mu$m (mean 180.1, median 45.2 mJy).The differences between mean and median values suggest a skewed distribution, with a subset of particularly bright sources increasing the average intensity consistent with a population containing both embedded protostars and more evolved, actively accreting clumps. 

Sources in UV-dominant the cluster show generally weaker molecular line intensities and IR fluxes with the smaller gap between mean and median values, which implies a more homogeneous population.  Mid-infrared fluxes remain substantial (8.0 $\mu$m: mean 179.5, median 16.1 mJy; 24 $\mu$m: mean 150.2, median 18.1 mJy).

The Prestellar cluster exhibits generally weak integrated molecular line intensities and C$_2$H and \ce{H^13CO^+} emission is below the detection threshold. The infrared emission in this cluster is also faint, for example, fluxes of at 8.0 $\mu$m: mean 54.6~mJy, median 3.07~mJy and 24~$\mu$m: mean~76.6 mJy, median 1.97~mJy, significantly lower than in the Active SF and UV-dominant clusters. This weak mid-infrared signature further supports the interpretation that these clumps are in an early, prestellar phase, prior to the formation of luminous embedded sources or significant dust heating.

Clusters include objects with different IR-based classifications, likely reflecting the fact that different tracers evolve on different timescales and that individual star-forming regions may host gas at multiple evolutionary stages~\citep[see][]{2025A&A...696A.202S, 2024A&A...689A..74A}, which reduces the clarity of feature-based interpretations and complicates simple classification schemes. Nevertheless, the three identified clusters, Prestellar, Active SF, and UV-dominant, appear to trace a coherent evolutionary sequence of molecular clumps. The Prestellar cluster represents cold, dense regions prior to star formation. The Active SF cluster is characterised by strong dense-gas emission and intense, but uneven mid-infrared output, meaning that there are some sources with exceptionally high IR brightness. The  UV-dominant cluster shows weaker, more uniform molecular emission and more evenly distributed infrared brightness. This trend reflects the physical transformation of molecular material as star formation advances and feedback reshapes the surrounding environment.

\begin{table}
    \centering
\begin{tabular}{llrr}
\hline
Cluster & Feature & Mean & Median  \\
\hline
\multirow{10}{*}{Active SF}&      
    C$_{2}$H     & 3.21  & 2.25   \\
&H$^{13}$CO$^{+}$ & 1.22  & 0.91   \\
& N$_{2}$H$^{+}$  & 8.41  & 6.05   \\
&      HCN        & 10.2  & 6.14   \\
&      HNC        & 6.43  & 4.21   \\
&   HCO$^{+}$     & 9.12  & 5.98   \\
&     8.0$\mu$m      &457.59 &28.06   \\
&      24$\mu$m      &180.1  &45.21   \\
&     4.5$\mu$m      &11.26  & 0.7    \\
&     3.6$\mu$m     &  38.7 &  4.02  \\

\hline
\multirow{10}{*}{UV-dominant}& 
       C$_{2}$H     &  1.9   &  1.43   \\
&  H$^{13}$CO$^{+}$ &  0.0   &  0.0   \\
&   N$_{2}$H$^{+}$  &  3.64  &  2.52   \\
&        HCN        &  6.29  &  3.82  \\
&        HNC        &  3.69  &  2.37  \\
&     HCO$^{+}$     &  5.17  &  3.5   \\
&       8.0$\mu$m     & 179.47 & 16.06  \\
&        24$\mu$m     & 150.16 & 18.07  \\
&       4.5$\mu$m     &  6.57  &  0.48  \\
&       3.6$\mu$m     & 20.37  &  2.74  \\

\hline
\multirow{10}{*}{Prestellar}&       
      C$_{2}$H     &  0.0  &  0.0   \\
& H$^{13}$CO$^{+}$ &  0.0  &  0.0   \\
&  N$_{2}$H$^{+}$  &  3.3  &  2.25  \\
&       HCN        & 7.84  &  2.65  \\
&       HNC        & 3.73  &  2.02  \\
&    HCO$^{+}$     & 6.03  &  2.82  \\
&      8.0$\mu$m       & 54.62 &  3.07  \\
&       24$\mu$m       & 76.63 &  1.97  \\
&      4.5$\mu$m       & 3.79  &  0.41  \\
&      3.6$\mu$m       & 12.56 &  2.05  \\
\hline
\end{tabular}
    \caption{Mean and median values of integrated molecular line intensities (K km s$^{-1}$) and infrared fluxes (mJy) for molecular clumps associated with the Prestellar, Active SF, and UV-dominant clusters identified in this work.}
    \label{tab:cluster_stat}
\end{table}

An additional noteworthy trend is seen in the ratio of mean values of HCN(1–0) to HNC(1–0) integrated intensities.  \citet{2013ApJ...777..157H} used MALT90 data of only 333 objects that were reliably classified manually as quiescent, protostellar or \HII{}/PDR. They showed that integrated intensities ratio of  HCN to HNC gradually increases as molecular clumps evolve, with median values of 1.07, 1.19, and 1.64 for quiescent, protostellar, and \HII{}/PDR sources, respectively. In our analysis (Table~\ref{tab:cluster_stat} ), the median ratios for the three clusters follow a similar progression(1.31, 1.46, and 1.61). \cite{2015A&A...576A.131H} performed mapping observations towards 38 high-mass star-forming regions of different evolutionary stages and showed that the global integrated intensity ratio HCN/HNC increase with evolutionary stage of sources (from 0.7 for quiescent sources up to 2.93 for \HII{} regions), further supporting the interpretation that this ratio traces chemical and physical evolution within the clumps. The systematic increase in HCN/HNC reflects the temperature-dependent conversion of HNC into HCN, which becomes more efficient in warmer gas. 

Interestingly, when we included dust temperature (derived in \cite{2015ApJ...815..130G}) as an additional feature in the analysis, its contribution to the clustering and classification appeared comparatively minor. This suggests that the algorithm identified an underlying link between the integrated intensities of the HCN/HNC ratio and dust temperature. At the same time, the results of \citet{Hacar2020} showed a correlation between the gas kinetic temperature derived from the HCN/HNC ratio and the dust temperature obtained from \textit{Herschel} observations (see their Figure~8c). Taken together the chemical tracers themselves, specifically HCN/HNC,  a tracer of the gas kinetic temperature, effectively reflect the temperature variations that differentiate evolutionary stages of molecular clumps. It is worth noting that HCN and HNC remained more important features than dust temperature in the analysis. We propose that molecular line intensities are sensitive to both the environment density and chemical evolution of the gas, which change significantly during star formation, whereas dust temperature represents only an averaged thermal state of the region. Consequently, it provides a less direct probe of the local physical and chemical conditions within the clumps.

\subsection{Supervised learning} \label{sec:supervised}

For both supervised learning methods, infrared features played a crucial role in defining the resulting categories. We propose that this may occur because  \cite{Rathborne2016} previously relied on IR features to establish their category definitions. In other words, the apparent dominance of IR fluxes in the supervised models may reflect the inheritance of feature relevance from the labelling scheme itself, rather than an intrinsic predictive superiority of those features. 

This behaviour is expected in a supervised framework and contrasts with the role of IR features in the unsupervised clustering analysis. The clustering analysis is fully unsupervised and does not use any category information. Clusters are identified based on how sources group together in the multidimensional feature space. Molecular line intensities introduce stronger source-to-source contrasts, which are more effective at separating objects into distinct density groups. IR fluxes, although physically important for tracing star formation, vary more smoothly across the data set and therefore contribute less to defining cluster boundaries. 

When we used the IR fluxes alone the resulted cross-validation score was around 0.59, which is almost the same score we achieved using IR fluxes along with values of integrated molecular line intensities.  Models based solely on integrated molecular line intensities achieved modest performance (cross-validation score was around 0.4). Despite the different cross-validation scores, in all classification attempts IR-dark regions consistently dominated among the identified clumps.

A noteworthy aspect of the analysis is that the cross-validation scores for both Random Forest and Gradient Boosting were relatively modest, consistently around 0.6. Several factors likely contribute to this outcome. First, key discriminative features may be missing from the data set. Although molecular line intensities and IR fluxes reflect important aspects of the  physical characteristics of the clumps, the absence of additional relevant features may constrain the predictive power of the models. In attempting to increase cross-validation scores, we consequently included a wide range of physical and chemical tracers as potential features, including \ce{CH_3OH} Class~II maser emission flags, dust temperature, integrated intensities of all the molecular lines presented in MALT90 survey and their ratios. Although these quantities reflect important aspects of the physical conditions within the clumps, the IR fluxes consistently emerged as the most influential predictors in all supervised learning experiments. This suggests that the absence of additional, complementary features, such as more detailed kinematic, chemical, or environmental parameters, may limit the overall predictive power of the models. It is also important to note that the evolutionary sequence of dense clumps represents a continuous process, rather than a set of sharply defined stages. Consequently, it is not possible to draw strict boundaries between evolutionary categories, as there will always be sources with intermediate or uncertain characteristics. Some degree of mixing between classes is therefore inevitable, and the best that can be achieved is to estimate the most probable evolutionary stage of each clump based on the available data.

\section{Conclusions}\label{sec:conclusions}

In this work, we demonstrated that machine learning methods are able to provide an objective and data-driven framework for classifying the evolutionary stages of molecular clumps. Our main conclusions are the following:
\begin{enumerate}
    \item In unsupervised learning, using only emission of five molecular lines (HCO$^+$, HNC, N$_2$H$^+$, HCN, and C$_2$H) data from the MALT90 survey, complemented by 870~$\mu$m continuum emission, the analysis revealed two distinct clusters: Prestellar and Active SF, dominated by IR-quiescent and protostellar objects and \HII{} regions, respectively. The most influential features driving this clustering were C$_2$H and N$_2$H$^+$, while the 870 $\mu$m continuum emission showed no significant effect on the cluster formation.
    \item With the inclusion of molecular emission of H$^{13}$CO$^+$  and \textit{Spitzer} IR fluxes the analysis revealed three stable clusters: Prestellar, Active SF, and UV-dominant, dominated by IR-quiescent, protostellar, and \HII{}+PDRs clumps, respectively, and primarily driven by variations in C$_2$H, N$_2$H$^+$, and H$^{13}$CO$^+$ emission. Even though H$^{13}$CO$^+$ emission itself is absent  within the UV-dominant cluster, its absence, combined with the presence of UV-tracing molecules such as C$_2$H, provides a chemical contrast that enables the algorithm to clearly isolate this group. The addition of \textit{Spitzer}  IR fluxes, however, did not alter the clustering structure.
    \item Including additional, less common molecular transitions from the MALT90 survey resulted only in further fragmentation of the existing structure. Nevertheless, the three principal clusters identified above consistently reappeared across all configurations, remaining the most stable and populated groups, albeit slightly reduced in size due to this fragmentation.
    \item Incorporating CH$_3$OH Class~II maser detection flags produced an additional cluster predominantly composed of protostellar and \HII{} region sources, which we interpret as a cluster with undergoing active massive star formation.
    \item In the supervised learning analysis, Random Forest and Gradient Boosting algorithms were employed to assign evolutionary categories to molecular clumps with previously uncertain classifications from the MALT90 survey. Using previously made manual classification which may contain hidden systematic and random errors, the models achieved moderate but consistent performance (cross-validation scores of approximately 0.6), indicating that the algorithms captured meaningful patterns in the data. The comparable accuracy of both methods suggests that the results are robust and not driven by algorithm-specific biases. The moderate classification accuracy likely reflects the absence of key discriminating features within the data set. The most influential predictors were infrared fluxes, consistent with the dependence of the training labels on IR-based classification schemes. Nevertheless, the models provided the most likely classifications for sources with previously uncertain labels, assigning the majority (436 out of 774 objects) of them to the IR-quiescent molecular clumps.

\end{enumerate}

In summary, our results demonstrate that integrated intensities of molecules at 90~GHz alone, even without complete photometric information, encode significant evolutionary information about star-forming clumps. However, not all tracers show clear or monotonic trends across evolutionary stages, which indicates that the chemical and physical evolution of molecular clumps proceeds uneven, with different tracers responding on distinct timescales to environmental changes. Such complexity complicates the definition of simple, universal classification schemes based on individual features. 

Our machine learning framework integrates molecular and photometric measurements, allowing the discovery of non-linear relationships between features and evolutionary stages. This enables a more objective and reproducible categorisation of molecular clumps, reducing the dependence on subjective thresholds.

Although our work focuses on MALT90 sources, the proposed methodology is broadly applicable to other surveys combining molecular and continuum data (e.g., Hi-GAL, JCMT, ALMA and so on). Expanding the feature space to include additional tracers, such as deuterated species, CO isotopologues, or radio continuum measurements, would likely refine the evolutionary mapping and better capture the interplay between chemistry, kinematics, and feedback processes. 

\section*{Acknowledgements}

We thank D.~A.~Ladeyschikov for fruitful discussions. We are grateful to A.~Guzm{\'a}n for information regarding the IR data. We are also sincerely thankful to the anonymous referee for their careful reading and constructive suggestions. The study is supported by the INASAN State Assignment Contract 'Voskhod' 124021200002-4. A.~B.~Ostrovskii is supported by the State Assignment Contract FEUZ-2025-003 (part of the Sec.~\ref{sec:methods}).

\section*{Data Availability}

The IR fluxes derived in this work as well as scripts used for analysis will be shared upon reasonable request to the corresponding author.



\bibliographystyle{mnras}
\bibliography{lib} 


\appendix

\section{Classification of previously uncertain clumps}





\onecolumn
\begin{longtable}{llrrllll}
\caption{Classification results for previously uncertain clumps with high confidence in both methods (probability $> 0.5$). The columns are as follows: object number; object name as listed in the MALT90 survey (with "S" indicating clumps where no second spectral component was identified, and "A"/"B" denoting multiple velocity components, where "A" corresponds to the brightest); galactic  longitude (l) and latitude (b); classification assigned by the Random Forest (RF Class) model along with its associated probability (RF Prob); and classification assigned by the Gradient Boosting (GB Class) model with its probability (GB Prob). Objects marked with asterisks indicate discrepancies in classification. \label{tab:classification_results}} \\
\hline
Num 
&Object & l & b & RF Class & RF Prob & GB Class & GB Prob \\
\hline
\endfirsthead

\multicolumn{8}{c}{\tablename~\thetable\ -- continued from previous page} \\
\hline
&Object & l & b & RF Class & RF Prob & GB Class & GB Prob \\
\hline
\endhead

\hline \multicolumn{8}{r}{{Continued on next page}} \\ \hline
\endfoot

\hline
\endlastfoot
1 & AGAL000.021-00.051\_A & 0.02 & -0.05 & Q & 0.60 & Q & 0.93 \\
2 & AGAL000.084+00.164\_S & 0.08 & 0.16 & Q & 0.80 & Q & 0.74 \\
3 & AGAL000.273-00.064\_A & 0.27 & -0.06 & Q & 0.57 & Q & 0.94 \\
4 & AGAL000.293-00.112\_B & 0.29 & -0.11 & P & 0.55 & P & 0.77 \\
5 & AGAL000.314-00.099\_A & 0.31 & -0.10 & Q & 0.54 & Q & 0.93 \\
6 & AGAL000.314-00.099\_B & 0.31 & -0.10 & P & 0.52 & P & 0.62 \\
7 & AGAL000.428-00.184\_A & 0.43 & -0.18 & Q & 0.86 & Q & 0.77 \\
8 & AGAL000.469+00.026\_S & 0.47 & 0.03 & Q & 0.97 & Q & 0.95 \\
9 & AGAL000.478-00.729\_S & 0.48 & -0.73 & Q & 0.63 & Q & 0.79 \\
10 & AGAL000.494-00.664\_S & 0.49 & -0.66 & A & 0.78 & A & 0.88 \\
11 & AGAL000.504+00.141\_S & 0.50 & 0.14 & Q & 0.58 & Q & 0.70 \\
12 & AGAL000.541+00.117\_A & 0.54 & 0.12 & Q & 0.69 & Q & 0.67 \\
13 & AGAL000.541+00.117\_B & 0.54 & 0.12 & Q & 0.76 & Q & 0.77 \\
14 & AGAL000.553-00.679\_S & 0.55 & -0.68 & Q & 0.79 & Q & 0.69 \\
15 & AGAL000.569-00.889\_S & 0.57 & -0.89 & Q & 0.73 & Q & 0.78 \\
16 & AGAL000.646+00.029\_A & 0.65 & 0.03 & Q & 0.98 & Q & 0.95 \\
17 & AGAL000.646+00.029\_B & 0.65 & 0.03 & Q & 0.91 & Q & 0.92 \\
18 & AGAL000.704-00.181\_S & 0.70 & -0.18 & Q & 0.95 & Q & 0.95 \\
19 & AGAL000.756-00.074\_S & 0.76 & -0.07 & Q & 0.98 & Q & 0.97 \\
20 & AGAL000.771-00.186\_S & 0.77 & -0.19 & Q & 0.98 & Q & 0.97 \\
21 & AGAL000.784-00.206\_A & 0.78 & -0.21 & Q & 0.98 & Q & 0.96 \\
22 & AGAL000.784-00.206\_B & 0.78 & -0.21 & Q & 0.88 & Q & 0.92 \\
23 & AGAL000.844-00.204\_A & 0.84 & -0.20 & Q & 0.87 & Q & 0.94 \\
24 & AGAL000.844-00.204\_B & 0.84 & -0.20 & Q & 0.66 & Q & 0.82 \\
25 & AGAL000.848-00.029\_A & 0.85 & -0.03 & Q & 0.88 & Q & 0.96 \\
26 & AGAL000.848-00.029\_B & 0.85 & -0.03 & Q & 0.85 & Q & 0.89 \\
27 & AGAL000.866-00.036\_S & 0.87 & -0.04 & Q & 0.82 & Q & 0.87 \\
28 & AGAL000.874-00.012\_A & 0.87 & -0.01 & Q & 0.76 & Q & 0.90 \\
29 & AGAL000.874-00.012\_B & 0.87 & -0.01 & Q & 0.79 & Q & 0.92 \\
30 & AGAL000.883+00.166\_S & 0.88 & 0.17 & Q & 0.98 & Q & 0.96 \\
31 & AGAL000.891+00.142\_S & 0.89 & 0.14 & Q & 0.98 & Q & 0.96 \\
32 & AGAL000.918-00.326\_S & 0.92 & -0.33 & Q & 0.58 & Q & 0.77 \\
33 & AGAL000.928-00.337\_S & 0.93 & -0.34 & Q & 0.76 & Q & 0.83 \\
34 & AGAL000.934-00.327\_S & 0.93 & -0.33 & Q & 0.69 & Q & 0.75 \\
35 & AGAL000.969-00.336\_S & 0.97 & -0.34 & Q & 0.80 & Q & 0.79 \\
36 & AGAL001.036-00.076\_A & 1.04 & -0.08 & Q & 0.95 & Q & 0.94 \\
37 & AGAL001.036-00.076\_B & 1.04 & -0.08 & Q & 0.51 & Q & 0.53 \\
38 & AGAL001.063-00.077\_A & 1.06 & -0.08 & Q & 0.59 & Q & 0.60 \\
39 & AGAL001.063-00.077\_B & 1.06 & -0.08 & Q & 0.81 & Q & 0.86 \\
40 & AGAL001.199+00.076\_S & 1.20 & 0.08 & Q & 0.98 & Q & 0.94 \\
41 & AGAL001.266+00.056\_A & 1.27 & 0.06 & Q & 0.97 & Q & 0.95 \\
42 & AGAL001.266+00.056\_B & 1.27 & 0.06 & Q & 0.70 & Q & 0.77 \\
43 & AGAL001.278-00.151\_A & 1.28 & -0.15 & Q & 0.99 & Q & 0.95 \\
44 & AGAL001.278-00.151\_B & 1.28 & -0.15 & Q & 0.70 & Q & 0.74 \\
45 & AGAL001.303-00.062\_A & 1.30 & -0.06 & Q & 0.97 & Q & 0.96 \\
46 & AGAL001.303-00.062\_B & 1.30 & -0.06 & Q & 0.74 & Q & 0.82 \\
47 & AGAL001.318+00.266\_S & 1.32 & 0.27 & Q & 0.96 & Q & 0.95 \\
48 & AGAL001.321-00.071\_S & 1.32 & -0.07 & Q & 0.94 & Q & 0.93 \\
49 & AGAL001.343-00.082\_A & 1.34 & -0.08 & Q & 0.96 & Q & 0.95 \\
50 & AGAL001.343-00.082\_B & 1.34 & -0.08 & Q & 0.64 & Q & 0.73 \\
51 & AGAL001.359+00.101\_S & 1.36 & 0.10 & Q & 0.89 & Q & 0.94 \\
52 & AGAL001.359+00.127\_A & 1.36 & 0.13 & Q & 0.65 & Q & 0.92 \\
53 & AGAL001.361+00.216\_S & 1.36 & 0.22 & Q & 0.93 & Q & 0.95 \\
54 & AGAL001.381+00.199\_S & 1.38 & 0.20 & Q & 0.99 & Q & 0.97 \\
55 & AGAL001.419-00.237\_S & 1.42 & -0.24 & Q & 0.86 & Q & 0.92 \\
56 & AGAL001.446+00.067\_S & 1.45 & 0.07 & Q & 0.94 & Q & 0.95 \\
57 & AGAL001.508-00.247\_A & 1.51 & -0.25 & Q & 0.79 & Q & 0.78 \\
58 & AGAL001.508-00.247\_B & 1.51 & -0.25 & Q & 0.72 & Q & 0.75 \\
59 & AGAL001.539-00.206\_A & 1.54 & -0.21 & Q & 0.93 & Q & 0.90 \\
60 & AGAL001.539-00.206\_B & 1.54 & -0.21 & Q & 0.83 & Q & 0.79 \\
61 & AGAL001.544-00.274\_A & 1.54 & -0.27 & Q & 0.87 & Q & 0.95 \\
62 & AGAL001.544-00.274\_B & 1.54 & -0.27 & Q & 0.66 & Q & 0.83 \\
63 & AGAL001.642-00.034\_A & 1.64 & -0.03 & Q & 0.86 & Q & 0.92 \\
64 & AGAL001.642-00.034\_B & 1.64 & -0.03 & Q & 0.83 & Q & 0.80 \\
65 & AGAL001.662-00.174\_S & 1.66 & -0.17 & Q & 0.86 & Q & 0.96 \\
66 & AGAL001.674-00.127\_S & 1.67 & -0.13 & Q & 0.90 & Q & 0.93 \\
67 & AGAL001.684-00.379\_A & 1.68 & -0.38 & Q & 0.81 & Q & 0.95 \\
68 & AGAL001.684-00.379\_B & 1.68 & -0.38 & Q & 0.84 & Q & 0.89 \\
69 & AGAL001.716-00.056\_S & 1.72 & -0.06 & Q & 0.98 & Q & 0.96 \\
70 & AGAL001.717-00.336\_A & 1.72 & -0.34 & Q & 0.89 & Q & 0.94 \\
71 & AGAL001.717-00.336\_B & 1.72 & -0.34 & Q & 0.79 & Q & 0.79 \\
72 & AGAL001.749-00.102\_S & 1.75 & -0.10 & Q & 0.94 & Q & 0.94 \\
73 & AGAL001.796-00.126\_S & 1.80 & -0.13 & Q & 0.86 & Q & 0.96 \\
74 & AGAL001.891-00.071\_S & 1.89 & -0.07 & Q & 0.90 & Q & 0.95 \\
75 & AGAL001.907-00.059\_S & 1.91 & -0.06 & Q & 0.88 & Q & 0.90 \\
76 & AGAL002.866+00.067\_S & 2.87 & 0.07 & Q & 0.94 & Q & 0.93 \\
77 & AGAL002.899+00.066\_S & 2.90 & 0.07 & Q & 0.90 & Q & 0.97 \\
78 & AGAL003.021-00.067\_A & 3.02 & -0.07 & Q & 0.97 & Q & 0.97 \\
79 & AGAL003.021-00.067\_B & 3.02 & -0.07 & Q & 0.96 & Q & 0.94 \\
80 & AGAL003.029-00.072\_A & 3.03 & -0.07 & Q & 0.85 & Q & 0.80 \\
81 & AGAL003.029-00.072\_B & 3.03 & -0.07 & Q & 0.79 & Q & 0.87 \\
82 & AGAL003.038-00.057\_A & 3.04 & -0.06 & Q & 0.94 & Q & 0.93 \\
83 & AGAL003.038-00.057\_B & 3.04 & -0.06 & Q & 0.85 & Q & 0.86 \\
84 & AGAL003.039+00.144\_A & 3.04 & 0.14 & Q & 0.87 & Q & 0.93 \\
85 & AGAL003.039+00.144\_B & 3.04 & 0.14 & Q & 0.88 & Q & 0.88 \\
86 & AGAL003.049+00.392\_S & 3.05 & 0.39 & Q & 0.89 & Q & 0.81 \\
87 & AGAL003.056+00.151\_A & 3.06 & 0.15 & Q & 0.86 & Q & 0.87 \\
88 & AGAL003.056+00.151\_B & 3.06 & 0.15 & Q & 0.91 & Q & 0.91 \\
89 & AGAL003.059-00.077\_A & 3.06 & -0.08 & Q & 0.94 & Q & 0.95 \\
90 & AGAL003.059-00.077\_B & 3.06 & -0.08 & Q & 0.76 & Q & 0.71 \\
91 & AGAL003.093+00.422\_S & 3.09 & 0.42 & Q & 0.92 & Q & 0.94 \\
92 & AGAL003.101+00.294\_A & 3.10 & 0.29 & Q & 0.85 & Q & 0.92 \\
93 & AGAL003.101+00.294\_B & 3.10 & 0.29 & Q & 0.70 & Q & 0.72 \\
94 & AGAL003.121+00.306\_S & 3.12 & 0.31 & Q & 0.74 & Q & 0.86 \\
95 & AGAL003.121+00.416\_S & 3.12 & 0.42 & Q & 0.94 & Q & 0.96 \\
96 & AGAL003.131+00.426\_S & 3.13 & 0.43 & Q & 0.85 & Q & 0.90 \\
97 & AGAL003.141+00.427\_A & 3.14 & 0.43 & Q & 0.85 & Q & 0.87 \\
98 & AGAL003.141+00.427\_B & 3.14 & 0.43 & Q & 0.92 & Q & 0.94 \\
99 & AGAL003.158+00.391\_A & 3.16 & 0.39 & Q & 0.84 & Q & 0.89 \\
100 & AGAL003.158+00.391\_B & 3.16 & 0.39 & Q & 0.81 & Q & 0.85 \\
101 & AGAL003.163+00.559\_A & 3.16 & 0.56 & Q & 0.95 & Q & 0.94 \\
102 & AGAL003.163+00.559\_B & 3.16 & 0.56 & Q & 0.82 & Q & 0.87 \\
103 & AGAL003.179+00.542\_A & 3.18 & 0.54 & Q & 0.94 & Q & 0.95 \\
104 & AGAL003.179+00.542\_B & 3.18 & 0.54 & Q & 0.81 & Q & 0.88 \\
105 & AGAL003.199+00.639\_S & 3.20 & 0.64 & Q & 0.95 & Q & 0.93 \\
106 & AGAL003.203+00.502\_S & 3.20 & 0.50 & Q & 0.92 & Q & 0.92 \\
107 & AGAL003.213+00.646\_S & 3.21 & 0.65 & Q & 0.90 & Q & 0.93 \\
108 & AGAL003.219+00.264\_S & 3.22 & 0.26 & Q & 0.84 & Q & 0.92 \\
109 & AGAL003.219+00.619\_S & 3.22 & 0.62 & Q & 0.88 & Q & 0.92 \\
110 & AGAL003.228+00.491\_A & 3.23 & 0.49 & Q & 0.94 & Q & 0.92 \\
111 & AGAL003.228+00.491\_B & 3.23 & 0.49 & Q & 0.81 & Q & 0.90 \\
112 & AGAL003.233+00.264\_A & 3.23 & 0.26 & Q & 0.94 & Q & 0.92 \\
113 & AGAL003.233+00.264\_B & 3.23 & 0.26 & Q & 0.88 & Q & 0.87 \\
114 & AGAL003.264+00.482\_S & 3.26 & 0.48 & Q & 0.92 & Q & 0.97 \\
115 & AGAL003.271+00.444\_A & 3.27 & 0.44 & Q & 0.98 & Q & 0.97 \\
116 & AGAL003.271+00.444\_B & 3.27 & 0.44 & Q & 0.75 & Q & 0.86 \\
117 & AGAL003.274+00.581\_S & 3.27 & 0.58 & Q & 0.91 & Q & 0.91 \\
118 & AGAL003.283+00.591\_S & 3.28 & 0.59 & Q & 0.82 & Q & 0.83 \\
119 & AGAL003.288+00.454\_S & 3.29 & 0.45 & Q & 0.95 & Q & 0.97 \\
120 & AGAL003.289+00.574\_S & 3.29 & 0.57 & Q & 0.86 & Q & 0.80 \\
121 & AGAL003.306+00.346\_S & 3.31 & 0.35 & Q & 0.65 & Q & 0.67 \\
122 & AGAL003.326+00.441\_A & 3.33 & 0.44 & Q & 0.93 & Q & 0.94 \\
123 & AGAL003.326+00.441\_B & 3.33 & 0.44 & Q & 0.77 & Q & 0.83 \\
124 & AGAL003.348+00.436\_A & 3.35 & 0.44 & Q & 0.97 & Q & 0.95 \\
125 & AGAL003.348+00.436\_B & 3.35 & 0.44 & Q & 0.74 & Q & 0.77 \\
126 & AGAL003.386+00.307\_S & 3.39 & 0.31 & Q & 0.86 & Q & 0.89 \\
127 & AGAL004.091+00.047\_S & 4.09 & 0.05 & Q & 0.72 & Q & 0.73 \\
128 & AGAL004.498-00.811\_S & 4.50 & -0.81 & Q & 0.70 & Q & 0.79 \\
129 & AGAL006.221-00.116\_S & 6.22 & -0.12 & Q & 0.66 & Q & 0.79 \\
130 & AGAL006.498-00.322\_S & 6.50 & -0.32 & Q & 0.74 & Q & 0.71 \\
131 & AGAL006.556-00.259\_S & 6.56 & -0.26 & Q & 0.89 & Q & 0.86 \\
132 & AGAL006.581-00.306\_S & 6.58 & -0.31 & Q & 0.75 & Q & 0.86 \\
133 & AGAL007.346-00.001\_S & 7.35 & -0.00 & Q & 0.78 & Q & 0.60 \\
134 & AGAL007.436-00.281\_S & 7.44 & -0.28 & Q & 0.77 & Q & 0.83 \\
135 & AGAL008.384-00.267\_S & 8.38 & -0.27 & Q & 0.55 & Q & 0.60 \\
136 & AGAL008.965-00.536\_S & 8.96 & -0.54 & Q & 0.75 & Q & 0.82 \\
137 & AGAL009.858-00.102\_S & 9.86 & -0.10 & Q & 0.67 & Q & 0.68 \\
138 & *AGAL010.051-00.211\_S & 10.05 & -0.21 & A & 0.59 & Q & 0.63 \\
139 & AGAL010.198-00.372\_S & 10.20 & -0.37 & A & 0.56 & A & 0.65 \\
140 & AGAL010.219-00.366\_S & 10.22 & -0.37 & Q & 0.53 & Q & 0.51 \\
141 & AGAL010.224-00.186\_S & 10.22 & -0.19 & Q & 0.76 & Q & 0.75 \\
142 & AGAL010.579-00.349\_S & 10.58 & -0.35 & Q & 0.84 & Q & 0.84 \\
143 & AGAL010.659-00.326\_S & 10.66 & -0.33 & Q & 0.53 & Q & 0.66 \\
144 & AGAL010.982-00.367\_S & 10.98 & -0.37 & Q & 0.66 & Q & 0.69 \\
145 & AGAL011.001-00.372\_S & 11.00 & -0.37 & Q & 0.91 & Q & 0.86 \\
146 & AGAL012.661-00.166\_A & 12.66 & -0.17 & Q & 0.76 & Q & 0.81 \\
147 & AGAL012.661-00.166\_B & 12.66 & -0.17 & Q & 0.77 & Q & 0.84 \\
148 & AGAL012.776-00.337\_A & 12.78 & -0.34 & Q & 0.86 & Q & 0.83 \\
149 & AGAL012.776-00.337\_B & 12.78 & -0.34 & Q & 0.90 & Q & 0.81 \\
150 & AGAL012.899-00.241\_S & 12.90 & -0.24 & A & 0.58 & A & 0.71 \\
151 & AGAL012.929-00.319\_S & 12.93 & -0.32 & Q & 0.51 & Q & 0.54 \\
152 & AGAL013.891-00.121\_A & 13.89 & -0.12 & Q & 0.60 & Q & 0.72 \\
153 & AGAL013.891-00.121\_B & 13.89 & -0.12 & Q & 0.58 & Q & 0.69 \\
154 & AGAL014.192-00.166\_S & 14.19 & -0.17 & A & 0.52 & A & 0.64 \\
155 & AGAL014.776-00.457\_S & 14.78 & -0.46 & Q & 0.86 & Q & 0.88 \\
156 & AGAL018.898-00.499\_S & 18.90 & -0.50 & A & 0.56 & A & 0.65 \\
157 & AGAL284.252-00.337\_S & 284.25 & -0.34 & Q & 0.51 & Q & 0.56 \\
158 & AGAL301.081+00.936\_S & 301.08 & 0.94 & Q & 0.70 & Q & 0.58 \\
159 & AGAL301.668-00.246\_S & 301.67 & -0.25 & Q & 0.92 & Q & 0.82 \\
160 & AGAL301.724+01.099\_S & 301.72 & 1.10 & Q & 0.74 & Q & 0.68 \\
161 & AGAL304.712+00.601\_S & 304.71 & 0.60 & Q & 0.70 & Q & 0.54 \\
162 & AGAL305.174+00.221\_S & 305.17 & 0.22 & Q & 0.64 & Q & 0.73 \\
163 & AGAL305.259+00.326\_S & 305.26 & 0.33 & Q & 0.50 & Q & 0.51 \\
164 & AGAL305.271-00.029\_S & 305.27 & -0.03 & P & 0.54 & P & 0.68 \\
165 & AGAL305.346+00.212\_S & 305.35 & 0.21 & P & 0.65 & P & 0.75 \\
166 & AGAL305.589+00.462\_S & 305.59 & 0.46 & Q & 0.58 & Q & 0.65 \\
167 & AGAL306.339-00.309\_S & 306.34 & -0.31 & Q & 0.60 & Q & 0.63 \\
168 & AGAL309.146-00.389\_S & 309.15 & -0.39 & Q & 0.74 & Q & 0.78 \\
169 & AGAL309.169-00.344\_S & 309.17 & -0.34 & Q & 0.57 & Q & 0.62 \\
170 & AGAL311.449+00.379\_A & 311.45 & 0.38 & Q & 0.86 & Q & 0.81 \\
171 & AGAL311.449+00.379\_B & 311.45 & 0.38 & Q & 0.75 & Q & 0.79 \\
172 & AGAL311.834+00.074\_S & 311.83 & 0.07 & Q & 0.76 & Q & 0.67 \\
173 & AGAL311.924+00.224\_S & 311.92 & 0.22 & P & 0.62 & P & 0.72 \\
174 & AGAL314.296+00.101\_S & 314.30 & 0.10 & A & 0.69 & A & 0.77 \\
175 & AGAL316.786-00.037\_S & 316.79 & -0.04 & A & 0.59 & A & 0.52 \\
176 & AGAL317.339+00.106\_S & 317.34 & 0.11 & Q & 0.57 & Q & 0.61 \\
177 & *AGAL317.441-00.374\_S & 317.44 & -0.37 & Q & 0.50 & A & 0.58 \\
178 & AGAL320.249+00.444\_S & 320.25 & 0.44 & H & 0.66 & H & 0.59 \\
179 & AGAL320.322-00.194\_S & 320.32 & -0.19 & P & 0.58 & P & 0.89 \\
180 & AGAL321.069-00.514\_S & 321.07 & -0.51 & P & 0.77 & P & 0.90 \\
181 & AGAL323.801+00.004\_S & 323.80 & 0.00 & Q & 0.85 & Q & 0.86 \\
182 & AGAL326.282-00.592\_S & 326.28 & -0.59 & Q & 0.72 & Q & 0.78 \\
183 & AGAL326.491+00.882\_S & 326.49 & 0.88 & Q & 0.59 & Q & 0.75 \\
184 & AGAL326.632+00.532\_S & 326.63 & 0.53 & P & 0.80 & P & 0.86 \\
185 & AGAL326.762+00.542\_S & 326.76 & 0.54 & Q & 0.81 & Q & 0.83 \\
186 & AGAL326.789-00.127\_S & 326.79 & -0.13 & Q & 0.68 & Q & 0.77 \\
187 & AGAL326.932-00.312\_S & 326.93 & -0.31 & Q & 0.62 & Q & 0.79 \\
188 & AGAL326.954-00.012\_S & 326.95 & -0.01 & A & 0.86 & A & 0.75 \\
189 & AGAL327.129+00.526\_S & 327.13 & 0.53 & Q & 0.88 & Q & 0.80 \\
190 & *AGAL327.216-00.494\_S & 327.22 & -0.49 & Q & 0.57 & A & 0.51 \\
191 & AGAL327.231-00.504\_S & 327.23 & -0.50 & P & 0.53 & P & 0.59 \\
192 & AGAL327.746-00.409\_S & 327.75 & -0.41 & Q & 0.59 & Q & 0.59 \\
193 & AGAL327.804-00.661\_S & 327.80 & -0.66 & Q & 0.84 & Q & 0.79 \\
194 & AGAL328.136+00.564\_S & 328.14 & 0.56 & Q & 0.60 & Q & 0.66 \\
195 & AGAL328.208-00.586\_S & 328.21 & -0.59 & P & 0.76 & P & 0.79 \\
196 & AGAL328.296-00.549\_S & 328.30 & -0.55 & A & 0.74 & A & 0.75 \\
197 & AGAL328.316-00.564\_S & 328.32 & -0.56 & Q & 0.55 & Q & 0.67 \\
198 & AGAL328.368-00.559\_S & 328.37 & -0.56 & A & 0.55 & A & 0.61 \\
199 & AGAL328.371-00.536\_S & 328.37 & -0.54 & Q & 0.63 & Q & 0.64 \\
200 & AGAL330.001+01.037\_A & 330.00 & 1.04 & Q & 0.80 & Q & 0.85 \\
201 & AGAL330.001+01.037\_B & 330.00 & 1.04 & Q & 0.84 & Q & 0.84 \\
202 & AGAL330.043+00.920\_S & 330.04 & 0.92 & Q & 0.64 & Q & 0.54 \\
203 & AGAL330.369+00.524\_S & 330.37 & 0.52 & Q & 0.52 & Q & 0.63 \\
204 & AGAL330.588+00.006\_S & 330.59 & 0.01 & Q & 0.59 & Q & 0.64 \\
205 & AGAL330.646-00.176\_S & 330.65 & -0.18 & Q & 0.65 & Q & 0.74 \\
206 & AGAL330.929-00.287\_S & 330.93 & -0.29 & A & 0.74 & A & 0.72 \\
207 & AGAL330.966-00.207\_S & 330.97 & -0.21 & Q & 0.86 & Q & 0.86 \\
208 & AGAL330.971-00.154\_S & 330.97 & -0.15 & Q & 0.56 & Q & 0.55 \\
209 & AGAL330.979-00.192\_S & 330.98 & -0.19 & Q & 0.63 & Q & 0.55 \\
210 & AGAL331.051-00.419\_S & 331.05 & -0.42 & Q & 0.68 & Q & 0.85 \\
211 & AGAL331.076-00.179\_S & 331.08 & -0.18 & Q & 0.58 & Q & 0.51 \\
212 & AGAL331.104-00.227\_S & 331.10 & -0.23 & Q & 0.82 & Q & 0.78 \\
213 & AGAL331.188-00.462\_S & 331.19 & -0.46 & Q & 0.57 & Q & 0.66 \\
214 & AGAL331.229-00.226\_A & 331.23 & -0.23 & Q & 0.65 & Q & 0.64 \\
215 & AGAL331.229-00.226\_B & 331.23 & -0.23 & Q & 0.60 & Q & 0.58 \\
216 & AGAL331.392-00.114\_S & 331.39 & -0.11 & Q & 0.77 & Q & 0.75 \\
217 & AGAL331.394-00.394\_S & 331.39 & -0.39 & Q & 0.79 & Q & 0.70 \\
218 & AGAL331.396-00.001\_S & 331.40 & -0.00 & Q & 0.80 & Q & 0.80 \\
219 & AGAL331.418-00.262\_S & 331.42 & -0.26 & Q & 0.65 & Q & 0.74 \\
220 & AGAL331.419-00.161\_S & 331.42 & -0.16 & Q & 0.69 & Q & 0.69 \\
221 & AGAL331.419-00.156\_S & 331.42 & -0.16 & Q & 0.86 & Q & 0.81 \\
222 & AGAL331.422-00.279\_S & 331.42 & -0.28 & Q & 0.65 & Q & 0.68 \\
223 & AGAL331.429-00.137\_S & 331.43 & -0.14 & Q & 0.76 & Q & 0.72 \\
224 & AGAL331.472-00.102\_S & 331.47 & -0.10 & Q & 0.67 & Q & 0.81 \\
225 & AGAL331.524-00.289\_A & 331.52 & -0.29 & Q & 0.70 & Q & 0.70 \\
226 & AGAL331.524-00.289\_B & 331.52 & -0.29 & Q & 0.77 & Q & 0.74 \\
227 & AGAL331.989-00.126\_S & 331.99 & -0.13 & Q & 0.94 & Q & 0.89 \\
228 & *AGAL332.112+00.926\_S & 332.11 & 0.93 & Q & 0.59 & A & 0.52 \\
229 & AGAL332.129+00.102\_A & 332.13 & 0.10 & Q & 0.73 & Q & 0.79 \\
230 & AGAL332.129+00.102\_B & 332.13 & 0.10 & Q & 0.66 & Q & 0.78 \\
231 & AGAL332.199+00.597\_S & 332.20 & 0.60 & Q & 0.66 & Q & 0.62 \\
232 & AGAL332.571-00.581\_S & 332.57 & -0.58 & Q & 0.82 & Q & 0.73 \\
233 & AGAL332.751-00.597\_S & 332.75 & -0.60 & P & 0.77 & P & 0.93 \\
234 & AGAL332.751-00.562\_S & 332.75 & -0.56 & P & 0.63 & P & 0.81 \\
235 & AGAL332.757-00.466\_S & 332.76 & -0.47 & Q & 0.80 & Q & 0.84 \\
236 & AGAL332.957-00.011\_S & 332.96 & -0.01 & Q & 0.85 & Q & 0.71 \\
237 & AGAL332.974-00.411\_S & 332.97 & -0.41 & P & 0.64 & P & 0.62 \\
238 & AGAL332.976-00.032\_S & 332.98 & -0.03 & Q & 0.76 & Q & 0.85 \\
239 & AGAL332.980-00.031\_S & 332.98 & -0.03 & Q & 0.85 & Q & 0.72 \\
240 & AGAL332.989-00.417\_S & 332.99 & -0.42 & P & 0.67 & P & 0.82 \\
241 & AGAL333.029-00.026\_S & 333.03 & -0.03 & P & 0.65 & P & 0.66 \\
242 & AGAL333.069+00.056\_S & 333.07 & 0.06 & Q & 0.60 & Q & 0.61 \\
243 & AGAL333.073-00.572\_S & 333.07 & -0.57 & Q & 0.79 & Q & 0.76 \\
244 & AGAL333.091-00.554\_S & 333.09 & -0.55 & Q & 0.56 & Q & 0.81 \\
245 & AGAL333.103-00.502\_S & 333.10 & -0.50 & Q & 0.50 & Q & 0.54 \\
246 & AGAL333.124+00.006\_S & 333.12 & 0.01 & Q & 0.94 & Q & 0.77 \\
247 & AGAL333.236+00.031\_A & 333.24 & 0.03 & A & 0.51 & A & 0.60 \\
248 & AGAL333.391+00.059\_A & 333.39 & 0.06 & Q & 0.92 & Q & 0.79 \\
249 & AGAL333.391+00.059\_B & 333.39 & 0.06 & Q & 0.93 & Q & 0.79 \\
250 & AGAL333.471-00.486\_S & 333.47 & -0.49 & A & 0.73 & A & 0.78 \\
251 & AGAL333.568+00.029\_S & 333.57 & 0.03 & A & 0.54 & A & 0.56 \\
252 & AGAL333.583+00.089\_S & 333.58 & 0.09 & Q & 0.80 & Q & 0.79 \\
253 & AGAL333.711-00.406\_S & 333.71 & -0.41 & Q & 0.71 & Q & 0.78 \\
254 & AGAL333.711-00.217\_S & 333.71 & -0.22 & Q & 0.64 & Q & 0.58 \\
255 & AGAL333.991+00.086\_S & 333.99 & 0.09 & Q & 0.62 & Q & 0.51 \\
256 & AGAL334.182+00.044\_S & 334.18 & 0.04 & Q & 0.69 & Q & 0.71 \\
257 & AGAL334.666+00.454\_S & 334.67 & 0.45 & Q & 0.72 & Q & 0.73 \\
258 & AGAL334.668+00.474\_S & 334.67 & 0.47 & Q & 0.80 & Q & 0.70 \\
259 & AGAL334.669+00.441\_S & 334.67 & 0.44 & Q & 0.52 & Q & 0.68 \\
260 & AGAL334.949-00.067\_S & 334.95 & -0.07 & Q & 0.64 & Q & 0.55 \\
261 & AGAL334.994-00.042\_S & 334.99 & -0.04 & Q & 0.79 & Q & 0.80 \\
262 & AGAL335.331+00.391\_S & 335.33 & 0.39 & Q & 0.80 & Q & 0.82 \\
263 & AGAL335.364+00.406\_S & 335.36 & 0.41 & Q & 0.87 & Q & 0.83 \\
264 & AGAL335.702-00.819\_S & 335.70 & -0.82 & Q & 0.55 & Q & 0.58 \\
265 & AGAL335.841-00.097\_S & 335.84 & -0.10 & Q & 0.91 & Q & 0.80 \\
266 & AGAL336.341-00.164\_S & 336.34 & -0.16 & Q & 0.84 & Q & 0.84 \\
267 & AGAL336.446-00.219\_S & 336.45 & -0.22 & Q & 0.79 & Q & 0.64 \\
268 & AGAL336.743+00.107\_S & 336.74 & 0.11 & P & 0.63 & P & 0.74 \\
269 & AGAL336.771+00.047\_S & 336.77 & 0.05 & P & 0.54 & P & 0.63 \\
270 & AGAL336.791+00.062\_S & 336.79 & 0.06 & H & 0.51 & H & 0.53 \\
271 & AGAL336.831+00.131\_S & 336.83 & 0.13 & Q & 0.68 & Q & 0.61 \\
272 & AGAL336.841+00.021\_S & 336.84 & 0.02 & P & 0.82 & P & 0.79 \\
273 & AGAL336.859+00.291\_S & 336.86 & 0.29 & Q & 0.90 & Q & 0.85 \\
274 & AGAL336.958-00.224\_S & 336.96 & -0.22 & Q & 0.78 & Q & 0.78 \\
275 & AGAL336.964-00.247\_S & 336.96 & -0.25 & Q & 0.90 & Q & 0.85 \\
276 & AGAL336.969-00.156\_S & 336.97 & -0.16 & P & 0.64 & P & 0.83 \\
277 & AGAL336.973+00.142\_S & 336.97 & 0.14 & Q & 0.59 & Q & 0.72 \\
278 & AGAL336.976-00.219\_S & 336.98 & -0.22 & Q & 0.56 & Q & 0.58 \\
279 & AGAL337.141-00.152\_S & 337.14 & -0.15 & Q & 0.60 & Q & 0.70 \\
280 & AGAL337.152-00.062\_S & 337.15 & -0.06 & Q & 0.81 & Q & 0.82 \\
281 & AGAL337.176+00.126\_S & 337.18 & 0.13 & Q & 0.69 & Q & 0.61 \\
282 & AGAL337.192-00.037\_S & 337.19 & -0.04 & Q & 0.75 & Q & 0.80 \\
283 & AGAL337.206-00.074\_S & 337.21 & -0.07 & Q & 0.74 & Q & 0.87 \\
284 & AGAL337.272-00.071\_S & 337.27 & -0.07 & Q & 0.80 & Q & 0.77 \\
285 & AGAL337.334-00.111\_S & 337.33 & -0.11 & Q & 0.70 & Q & 0.57 \\
286 & AGAL337.341-00.141\_S & 337.34 & -0.14 & Q & 0.82 & Q & 0.77 \\
287 & AGAL337.342-00.119\_S & 337.34 & -0.12 & Q & 0.79 & Q & 0.81 \\
288 & AGAL337.348-00.159\_S & 337.35 & -0.16 & Q & 0.73 & Q & 0.56 \\
289 & AGAL337.451-00.382\_S & 337.45 & -0.38 & A & 0.75 & A & 0.73 \\
290 & AGAL337.511+00.091\_S & 337.51 & 0.09 & Q & 0.73 & Q & 0.76 \\
291 & AGAL337.602-00.034\_S & 337.60 & -0.03 & A & 0.62 & A & 0.55 \\
292 & AGAL337.632+00.144\_S & 337.63 & 0.14 & Q & 0.68 & Q & 0.55 \\
293 & AGAL337.669-00.044\_S & 337.67 & -0.04 & Q & 0.71 & Q & 0.68 \\
294 & AGAL337.739+00.092\_S & 337.74 & 0.09 & Q & 0.74 & Q & 0.79 \\
295 & *AGAL337.794-00.002\_S & 337.79 & -0.00 & Q & 0.55 & A & 0.52 \\
296 & AGAL337.942-00.006\_S & 337.94 & -0.01 & H & 0.58 & H & 0.59 \\
297 & AGAL337.946+00.012\_S & 337.95 & 0.01 & Q & 0.68 & Q & 0.57 \\
298 & AGAL337.987+00.024\_S & 337.99 & 0.02 & Q & 0.51 & Q & 0.51 \\
299 & AGAL338.007-00.004\_A & 338.01 & -0.00 & Q & 0.51 & Q & 0.57 \\
300 & AGAL338.007-00.004\_B & 338.01 & -0.00 & Q & 0.50 & Q & 0.57 \\
301 & AGAL338.024-00.019\_S & 338.02 & -0.02 & Q & 0.86 & Q & 0.82 \\
302 & AGAL338.036-00.042\_S & 338.04 & -0.04 & Q & 0.69 & Q & 0.79 \\
303 & AGAL338.091-00.191\_S & 338.09 & -0.19 & Q & 0.65 & Q & 0.62 \\
304 & AGAL338.094-00.011\_S & 338.09 & -0.01 & Q & 0.88 & Q & 0.78 \\
305 & AGAL338.114+00.006\_S & 338.11 & 0.01 & Q & 0.54 & Q & 0.61 \\
306 & AGAL338.131+00.099\_S & 338.13 & 0.10 & Q & 0.73 & Q & 0.87 \\
307 & AGAL338.147+00.109\_S & 338.15 & 0.11 & Q & 0.77 & Q & 0.87 \\
308 & AGAL338.402+00.032\_S & 338.40 & 0.03 & A & 0.60 & A & 0.52 \\
309 & AGAL338.446-00.006\_A & 338.45 & -0.01 & P & 0.69 & P & 0.58 \\
310 & AGAL338.446+00.044\_A & 338.45 & 0.04 & A & 0.73 & A & 0.63 \\
311 & AGAL338.446+00.044\_B & 338.45 & 0.04 & A & 0.54 & A & 0.57 \\
312 & AGAL338.494+00.044\_S & 338.49 & 0.04 & Q & 0.57 & Q & 0.65 \\
313 & AGAL338.552+00.011\_A & 338.55 & 0.01 & Q & 0.64 & Q & 0.65 \\
314 & AGAL338.552+00.011\_B & 338.55 & 0.01 & Q & 0.58 & Q & 0.64 \\
315 & AGAL338.561-00.012\_A & 338.56 & -0.01 & Q & 0.77 & Q & 0.80 \\
316 & AGAL338.561-00.012\_B & 338.56 & -0.01 & Q & 0.79 & Q & 0.80 \\
317 & AGAL338.621+00.022\_S & 338.62 & 0.02 & Q & 0.70 & Q & 0.79 \\
318 & AGAL338.786+00.476\_S & 338.79 & 0.48 & Q & 0.85 & Q & 0.89 \\
319 & AGAL338.834+00.479\_S & 338.83 & 0.48 & Q & 0.80 & Q & 0.68 \\
320 & AGAL338.889+00.394\_A & 338.89 & 0.39 & Q & 0.86 & Q & 0.91 \\
321 & AGAL338.889+00.394\_B & 338.89 & 0.39 & Q & 0.89 & Q & 0.90 \\
322 & AGAL338.929-00.081\_S & 338.93 & -0.08 & P & 0.72 & P & 0.56 \\
323 & AGAL338.949+00.531\_S & 338.95 & 0.53 & Q & 0.76 & Q & 0.86 \\
324 & AGAL339.444+00.092\_S & 339.44 & 0.09 & Q & 0.68 & Q & 0.54 \\
325 & AGAL339.548-00.129\_S & 339.55 & -0.13 & Q & 0.59 & Q & 0.70 \\
326 & AGAL339.608-00.116\_S & 339.61 & -0.12 & Q & 0.58 & Q & 0.70 \\
327 & AGAL340.094-00.316\_S & 340.09 & -0.32 & A & 0.62 & A & 0.71 \\
328 & AGAL340.096-00.022\_S & 340.10 & -0.02 & Q & 0.60 & Q & 0.70 \\
329 & AGAL340.206-00.049\_S & 340.21 & -0.05 & Q & 0.68 & Q & 0.79 \\
330 & AGAL340.212-00.022\_S & 340.21 & -0.02 & P & 0.57 & P & 0.52 \\
331 & AGAL340.226-00.017\_S & 340.23 & -0.02 & Q & 0.64 & Q & 0.67 \\
332 & AGAL340.274-00.029\_S & 340.27 & -0.03 & Q & 0.64 & Q & 0.59 \\
333 & AGAL340.284-00.072\_S & 340.28 & -0.07 & Q & 0.63 & Q & 0.65 \\
334 & AGAL340.299-00.222\_S & 340.30 & -0.22 & A & 0.56 & A & 0.62 \\
335 & AGAL340.311+00.557\_S & 340.31 & 0.56 & Q & 0.62 & Q & 0.64 \\
336 & AGAL340.618-00.646\_S & 340.62 & -0.65 & Q & 0.91 & Q & 0.83 \\
337 & AGAL340.636-00.664\_S & 340.64 & -0.66 & Q & 0.75 & Q & 0.77 \\
338 & AGAL340.719-00.976\_S & 340.72 & -0.98 & Q & 0.74 & Q & 0.74 \\
339 & AGAL340.724-00.979\_S & 340.72 & -0.98 & Q & 0.76 & Q & 0.81 \\
340 & AGAL340.901-00.346\_S & 340.90 & -0.35 & Q & 0.72 & Q & 0.69 \\
341 & AGAL340.984-00.352\_S & 340.98 & -0.35 & Q & 0.86 & Q & 0.79 \\
342 & AGAL341.009-00.361\_S & 341.01 & -0.36 & Q & 0.61 & Q & 0.73 \\
343 & AGAL341.172-00.256\_S & 341.17 & -0.26 & Q & 0.87 & Q & 0.85 \\
344 & AGAL341.211-00.271\_S & 341.21 & -0.27 & Q & 0.68 & Q & 0.84 \\
345 & AGAL341.267-00.287\_S & 341.27 & -0.29 & Q & 0.62 & Q & 0.84 \\
346 & AGAL341.721+00.059\_S & 341.72 & 0.06 & Q & 0.85 & Q & 0.80 \\
347 & AGAL342.156+00.436\_S & 342.16 & 0.44 & Q & 0.92 & Q & 0.91 \\
348 & AGAL342.273+00.489\_A & 342.27 & 0.49 & Q & 0.70 & Q & 0.73 \\
349 & AGAL342.273+00.489\_B & 342.27 & 0.49 & Q & 0.62 & Q & 0.71 \\
350 & AGAL342.916-00.141\_S & 342.92 & -0.14 & Q & 0.58 & Q & 0.53 \\
351 & AGAL343.498+00.009\_S & 343.50 & 0.01 & P & 0.77 & P & 0.86 \\
352 & AGAL343.522-00.077\_S & 343.52 & -0.08 & Q & 0.68 & Q & 0.65 \\
353 & AGAL343.877+00.106\_S & 343.88 & 0.11 & Q & 0.88 & Q & 0.90 \\
354 & AGAL344.112-00.637\_S & 344.11 & -0.64 & Q & 0.52 & Q & 0.61 \\
355 & AGAL344.194-00.617\_S & 344.19 & -0.62 & Q & 0.78 & Q & 0.76 \\
356 & AGAL344.224-00.674\_S & 344.22 & -0.67 & Q & 0.76 & Q & 0.76 \\
357 & AGAL344.809-00.632\_S & 344.81 & -0.63 & Q & 0.64 & Q & 0.79 \\
358 & AGAL345.196-00.744\_S & 345.20 & -0.74 & A & 0.67 & A & 0.79 \\
359 & AGAL345.389-00.932\_S & 345.39 & -0.93 & P & 0.58 & P & 0.73 \\
360 & AGAL345.439-00.941\_S & 345.44 & -0.94 & H & 0.76 & H & 0.72 \\
361 & AGAL345.503-00.032\_S & 345.50 & -0.03 & Q & 0.59 & Q & 0.65 \\
362 & AGAL345.639-00.006\_S & 345.64 & -0.01 & Q & 0.83 & Q & 0.81 \\
363 & AGAL346.486+00.146\_S & 346.49 & 0.15 & Q & 0.86 & Q & 0.80 \\
364 & AGAL347.234+00.032\_S & 347.23 & 0.03 & P & 0.62 & P & 0.61 \\
365 & AGAL347.611+00.256\_S & 347.61 & 0.26 & Q & 0.62 & Q & 0.63 \\
366 & AGAL347.627+00.122\_S & 347.63 & 0.12 & Q & 0.58 & Q & 0.76 \\
367 & AGAL347.993-00.431\_S & 347.99 & -0.43 & Q & 0.76 & Q & 0.56 \\
368 & AGAL348.121+00.271\_S & 348.12 & 0.27 & Q & 0.64 & Q & 0.54 \\
369 & AGAL348.148+00.469\_S & 348.15 & 0.47 & Q & 0.60 & Q & 0.74 \\
370 & AGAL348.156+00.506\_S & 348.16 & 0.51 & Q & 0.68 & Q & 0.70 \\
371 & AGAL348.626-00.907\_S & 348.63 & -0.91 & A & 0.51 & A & 0.64 \\
372 & AGAL348.676-01.051\_S & 348.68 & -1.05 & P & 0.53 & P & 0.64 \\
373 & AGAL348.749-00.986\_S & 348.75 & -0.99 & A & 0.50 & A & 0.61 \\
374 & AGAL348.896+00.111\_S & 348.90 & 0.11 & Q & 0.76 & Q & 0.83 \\
375 & AGAL349.168+00.072\_S & 349.17 & 0.07 & Q & 0.78 & Q & 0.76 \\
376 & AGAL349.812-00.517\_S & 349.81 & -0.52 & P & 0.70 & P & 0.69 \\
377 & AGAL350.002-00.552\_S & 350.00 & -0.55 & Q & 0.79 & Q & 0.85 \\
378 & AGAL350.016-00.534\_S & 350.02 & -0.53 & Q & 0.69 & Q & 0.78 \\
379 & AGAL350.126+00.112\_S & 350.13 & 0.11 & Q & 0.74 & Q & 0.79 \\
380 & AGAL350.154+00.009\_S & 350.15 & 0.01 & Q & 0.88 & Q & 0.85 \\
381 & AGAL350.176+00.036\_S & 350.18 & 0.04 & Q & 0.72 & Q & 0.79 \\
382 & AGAL350.184+00.002\_S & 350.18 & 0.00 & Q & 0.63 & Q & 0.80 \\
383 & *AGAL350.201+00.036\_S & 350.20 & 0.04 & A & 0.51 & Q & 0.52 \\
384 & AGAL350.354+00.144\_S & 350.35 & 0.14 & Q & 0.73 & Q & 0.79 \\
385 & AGAL350.367+00.189\_S & 350.37 & 0.19 & Q & 0.64 & Q & 0.83 \\
386 & AGAL350.499-00.394\_S & 350.50 & -0.39 & Q & 0.77 & Q & 0.77 \\
387 & AGAL350.539-00.369\_S & 350.54 & -0.37 & Q & 0.76 & Q & 0.79 \\
388 & AGAL350.544+00.957\_S & 350.54 & 0.96 & Q & 0.55 & Q & 0.80 \\
389 & AGAL350.566+00.949\_S & 350.57 & 0.95 & Q & 0.76 & Q & 0.87 \\
390 & AGAL350.581+00.367\_S & 350.58 & 0.37 & Q & 0.68 & Q & 0.83 \\
391 & AGAL350.729+00.927\_S & 350.73 & 0.93 & Q & 0.72 & Q & 0.74 \\
392 & AGAL350.736+00.859\_S & 350.74 & 0.86 & Q & 0.57 & Q & 0.64 \\
393 & AGAL350.961+00.552\_S & 350.96 & 0.55 & A & 0.58 & A & 0.50 \\
394 & AGAL350.961+00.741\_S & 350.96 & 0.74 & Q & 0.71 & Q & 0.66 \\
395 & AGAL350.967+00.546\_S & 350.97 & 0.55 & Q & 0.54 & Q & 0.57 \\
396 & AGAL351.141+00.776\_S & 351.14 & 0.78 & P & 0.63 & P & 0.80 \\
397 & AGAL351.159+00.749\_S & 351.16 & 0.75 & P & 0.62 & P & 0.74 \\
398 & AGAL351.173+00.661\_S & 351.17 & 0.66 & P & 0.65 & P & 0.81 \\
399 & AGAL351.308+00.684\_S & 351.31 & 0.68 & P & 0.89 & P & 0.95 \\
400 & AGAL351.353+00.696\_S & 351.35 & 0.70 & P & 0.60 & P & 0.54 \\
401 & *AGAL351.409+00.567\_S & 351.41 & 0.57 & Q & 0.53 & P & 0.50 \\
402 & AGAL351.414+00.594\_S & 351.41 & 0.59 & P & 0.60 & P & 0.86 \\
403 & *AGAL351.421+00.551\_S & 351.42 & 0.55 & Q & 0.57 & P & 0.52 \\
404 & AGAL351.456+00.666\_S & 351.46 & 0.67 & Q & 0.73 & Q & 0.90 \\
405 & AGAL351.461+00.644\_S & 351.46 & 0.64 & Q & 0.77 & Q & 0.76 \\
406 & AGAL351.466+00.591\_S & 351.47 & 0.59 & Q & 0.67 & Q & 0.80 \\
407 & AGAL351.466+00.682\_S & 351.47 & 0.68 & Q & 0.59 & Q & 0.85 \\
408 & AGAL351.469+00.672\_S & 351.47 & 0.67 & Q & 0.58 & Q & 0.74 \\
409 & *AGAL351.491+00.691\_S & 351.49 & 0.69 & Q & 0.52 & P & 0.51 \\
410 & AGAL351.498+00.646\_S & 351.50 & 0.65 & Q & 0.57 & Q & 0.71 \\
411 & AGAL351.766+00.212\_S & 351.77 & 0.21 & Q & 0.78 & Q & 0.80 \\
412 & AGAL351.784+00.212\_S & 351.78 & 0.21 & Q & 0.70 & Q & 0.57 \\
413 & AGAL351.804+00.622\_S & 351.80 & 0.62 & Q & 0.64 & Q & 0.73 \\
414 & AGAL352.113+00.191\_S & 352.11 & 0.19 & Q & 0.83 & Q & 0.82 \\
415 & AGAL352.181-00.154\_S & 352.18 & -0.15 & Q & 0.86 & Q & 0.77 \\
416 & AGAL352.219-00.087\_S & 352.22 & -0.09 & Q & 0.58 & Q & 0.56 \\
417 & AGAL352.442-00.182\_S & 352.44 & -0.18 & Q & 0.71 & Q & 0.78 \\
418 & AGAL352.471+00.794\_S & 352.47 & 0.79 & Q & 0.83 & Q & 0.85 \\
419 & AGAL352.472-00.187\_A & 352.47 & -0.19 & Q & 0.77 & Q & 0.78 \\
420 & AGAL352.472-00.187\_B & 352.47 & -0.19 & Q & 0.72 & Q & 0.77 \\
421 & AGAL352.964+00.956\_S & 352.96 & 0.96 & A & 0.50 & A & 0.60 \\
422 & AGAL352.992+00.921\_S & 352.99 & 0.92 & P & 0.83 & P & 0.92 \\
423 & AGAL352.996+00.554\_S & 353.00 & 0.55 & Q & 0.64 & Q & 0.72 \\
424 & AGAL352.999+00.574\_S & 353.00 & 0.57 & P & 0.77 & P & 0.90 \\
425 & AGAL353.019+00.547\_S & 353.02 & 0.55 & Q & 0.64 & Q & 0.68 \\
426 & AGAL353.051+00.446\_S & 353.05 & 0.45 & P & 0.86 & P & 0.96 \\
427 & AGAL353.062+00.406\_S & 353.06 & 0.41 & P & 0.73 & P & 0.72 \\
428 & AGAL353.076+00.441\_S & 353.08 & 0.44 & Q & 0.59 & Q & 0.65 \\
429 & AGAL353.079+00.981\_S & 353.08 & 0.98 & Q & 0.53 & Q & 0.63 \\
430 & AGAL353.084+00.436\_S & 353.08 & 0.44 & A & 0.61 & A & 0.66 \\
431 & AGAL353.121+00.952\_S & 353.12 & 0.95 & P & 0.82 & P & 0.92 \\
432 & AGAL353.127+00.614\_S & 353.13 & 0.61 & P & 0.55 & P & 0.84 \\
433 & AGAL353.142+00.631\_S & 353.14 & 0.63 & P & 0.66 & P & 0.86 \\
434 & AGAL353.149+00.961\_S & 353.15 & 0.96 & P & 0.61 & P & 0.77 \\
435 & AGAL353.229+00.672\_S & 353.23 & 0.67 & P & 0.89 & P & 0.94 \\
436 & AGAL353.296+00.637\_S & 353.30 & 0.64 & P & 0.68 & P & 0.87 \\
437 & AGAL353.309+00.661\_S & 353.31 & 0.66 & P & 0.82 & P & 0.93 \\
438 & AGAL353.354-00.109\_A & 353.35 & -0.11 & Q & 0.81 & Q & 0.83 \\
439 & AGAL353.354-00.109\_B & 353.35 & -0.11 & Q & 0.82 & Q & 0.83 \\
440 & AGAL353.429-00.097\_A & 353.43 & -0.10 & Q & 0.81 & Q & 0.77 \\
441 & AGAL353.429-00.097\_B & 353.43 & -0.10 & Q & 0.77 & Q & 0.73 \\
442 & AGAL353.556+00.652\_S & 353.56 & 0.65 & Q & 0.73 & Q & 0.67 \\
443 & AGAL354.326+00.466\_S & 354.33 & 0.47 & Q & 0.83 & Q & 0.87 \\
444 & AGAL354.341+00.474\_S & 354.34 & 0.47 & Q & 0.80 & Q & 0.91 \\
445 & AGAL354.363+00.476\_S & 354.36 & 0.48 & Q & 0.83 & Q & 0.94 \\
446 & AGAL354.554+00.019\_S & 354.55 & 0.02 & Q & 0.63 & Q & 0.85 \\
447 & AGAL354.659+00.507\_S & 354.66 & 0.51 & Q & 0.77 & Q & 0.75 \\
448 & AGAL354.688+00.544\_S & 354.69 & 0.54 & Q & 0.71 & Q & 0.75 \\
449 & AGAL354.714+00.322\_B & 354.71 & 0.32 & Q & 0.67 & Q & 0.59 \\
450 & AGAL354.716+00.554\_S & 354.72 & 0.55 & Q & 0.72 & Q & 0.64 \\
451 & AGAL354.766+00.359\_A & 354.77 & 0.36 & Q & 0.62 & Q & 0.73 \\
452 & AGAL354.766+00.359\_B & 354.77 & 0.36 & Q & 0.73 & Q & 0.75 \\
453 & AGAL354.769+00.394\_S & 354.77 & 0.39 & Q & 0.73 & Q & 0.80 \\
454 & AGAL354.834+00.377\_A & 354.83 & 0.38 & Q & 0.87 & Q & 0.94 \\
455 & AGAL354.834+00.377\_B & 354.83 & 0.38 & Q & 0.85 & Q & 0.84 \\
456 & AGAL354.838+00.372\_S & 354.84 & 0.37 & Q & 0.79 & Q & 0.94 \\
457 & AGAL354.843+00.351\_A & 354.84 & 0.35 & Q & 0.63 & Q & 0.79 \\
458 & AGAL354.843+00.351\_B & 354.84 & 0.35 & Q & 0.63 & Q & 0.74 \\
459 & AGAL354.854+00.341\_S & 354.85 & 0.34 & Q & 0.63 & Q & 0.82 \\
460 & AGAL355.514-00.097\_S & 355.51 & -0.10 & Q & 0.64 & Q & 0.78 \\
461 & AGAL355.606-00.054\_S & 355.61 & -0.05 & Q & 0.84 & Q & 0.84 \\
462 & AGAL355.628-00.067\_S & 355.63 & -0.07 & Q & 0.82 & Q & 0.85 \\
463 & AGAL355.638-00.057\_S & 355.64 & -0.06 & Q & 0.78 & Q & 0.67 \\
464 & AGAL356.226+00.706\_S & 356.23 & 0.71 & Q & 0.60 & Q & 0.68 \\
465 & AGAL356.367+00.239\_S & 356.37 & 0.24 & Q & 0.84 & Q & 0.79 \\
466 & AGAL358.186-00.012\_S & 358.19 & -0.01 & Q & 0.79 & Q & 0.75 \\
467 & AGAL358.564-00.797\_S & 358.56 & -0.80 & Q & 0.79 & Q & 0.76 \\
468 & AGAL358.576-00.191\_S & 358.58 & -0.19 & Q & 0.61 & Q & 0.70 \\
469 & AGAL358.704-00.101\_S & 358.70 & -0.10 & Q & 0.60 & Q & 0.66 \\
470 & AGAL358.721-00.129\_A & 358.72 & -0.13 & Q & 0.54 & Q & 0.61 \\
471 & AGAL358.721-00.129\_B & 358.72 & -0.13 & Q & 0.70 & Q & 0.71 \\
472 & AGAL358.721-00.109\_S & 358.72 & -0.11 & Q & 0.55 & Q & 0.68 \\
473 & AGAL358.734-00.116\_S & 358.73 & -0.12 & Q & 0.71 & Q & 0.51 \\
474 & AGAL358.779-00.117\_A & 358.78 & -0.12 & Q & 0.85 & Q & 0.84 \\
475 & AGAL358.779-00.117\_B & 358.78 & -0.12 & Q & 0.90 & Q & 0.87 \\
476 & AGAL358.796-00.136\_A & 358.80 & -0.14 & Q & 0.89 & Q & 0.87 \\
477 & AGAL358.796-00.136\_B & 358.80 & -0.14 & Q & 0.93 & Q & 0.83 \\
478 & AGAL358.798-00.117\_A & 358.80 & -0.12 & Q & 0.71 & Q & 0.70 \\
479 & AGAL358.798-00.117\_B & 358.80 & -0.12 & Q & 0.87 & Q & 0.86 \\
480 & AGAL358.808-00.131\_A & 358.81 & -0.13 & Q & 0.92 & Q & 0.91 \\
481 & AGAL358.808-00.131\_B & 358.81 & -0.13 & Q & 0.84 & Q & 0.91 \\
482 & AGAL358.887-00.354\_S & 358.89 & -0.35 & Q & 0.78 & Q & 0.82 \\
483 & AGAL359.196+00.174\_A & 359.20 & 0.17 & Q & 0.91 & Q & 0.92 \\
484 & AGAL359.196+00.174\_B & 359.20 & 0.17 & Q & 0.86 & Q & 0.87 \\
485 & AGAL359.199+00.171\_S & 359.20 & 0.17 & Q & 0.76 & Q & 0.82 \\
486 & AGAL359.201-00.076\_A & 359.20 & -0.08 & Q & 0.90 & Q & 0.97 \\
487 & AGAL359.201-00.076\_B & 359.20 & -0.08 & Q & 0.77 & Q & 0.90 \\
488 & AGAL359.217+00.181\_A & 359.22 & 0.18 & Q & 0.70 & Q & 0.75 \\
489 & AGAL359.217+00.181\_B & 359.22 & 0.18 & Q & 0.56 & Q & 0.72 \\
490 & AGAL359.281-00.079\_A & 359.28 & -0.08 & Q & 0.83 & Q & 0.89 \\
491 & AGAL359.361-00.009\_S & 359.36 & -0.01 & Q & 0.80 & Q & 0.87 \\
492 & AGAL359.369-00.029\_A & 359.37 & -0.03 & Q & 0.77 & Q & 0.82 \\
493 & AGAL359.369-00.029\_B & 359.37 & -0.03 & Q & 0.53 & Q & 0.64 \\
494 & AGAL359.422+00.024\_S & 359.42 & 0.02 & Q & 0.57 & Q & 0.59 \\
495 & AGAL359.424-00.171\_A & 359.42 & -0.17 & Q & 0.80 & Q & 0.86 \\
496 & AGAL359.424-00.171\_B & 359.42 & -0.17 & Q & 0.78 & Q & 0.84 \\
497 & AGAL359.477+00.136\_S & 359.48 & 0.14 & Q & 0.74 & Q & 0.72 \\
498 & AGAL359.482+00.132\_S & 359.48 & 0.13 & Q & 0.96 & Q & 0.95 \\
499 & AGAL359.489-00.214\_A & 359.49 & -0.21 & Q & 0.88 & Q & 0.90 \\
500 & AGAL359.489-00.214\_B & 359.49 & -0.21 & Q & 0.80 & Q & 0.82 \\
501 & AGAL359.502+00.116\_S & 359.50 & 0.12 & Q & 0.74 & Q & 0.60 \\
502 & AGAL359.524+00.134\_S & 359.52 & 0.13 & Q & 0.81 & Q & 0.84 \\
503 & AGAL359.536-00.184\_S & 359.54 & -0.18 & Q & 0.91 & Q & 0.92 \\
504 & AGAL359.566-00.161\_A & 359.57 & -0.16 & Q & 0.86 & Q & 0.92 \\
505 & AGAL359.566-00.161\_B & 359.57 & -0.16 & Q & 0.65 & Q & 0.71 \\
506 & AGAL359.572-00.174\_A & 359.57 & -0.17 & Q & 0.78 & Q & 0.90 \\
507 & AGAL359.572-00.174\_B & 359.57 & -0.17 & Q & 0.74 & Q & 0.57 \\
508 & AGAL359.597+00.016\_A & 359.60 & 0.02 & Q & 0.62 & Q & 0.80 \\
509 & AGAL359.599-00.221\_A & 359.60 & -0.22 & Q & 0.52 & Q & 0.72 \\
510 & AGAL359.599-00.221\_B & 359.60 & -0.22 & Q & 0.60 & Q & 0.90 \\
511 & AGAL359.704-00.036\_A & 359.70 & -0.04 & Q & 0.68 & Q & 0.85 \\
512 & AGAL359.742+00.031\_A & 359.74 & 0.03 & Q & 0.66 & Q & 0.86 \\
513 & AGAL359.742+00.031\_B & 359.74 & 0.03 & Q & 0.89 & Q & 0.94 \\
514 & AGAL359.764-00.119\_S & 359.76 & -0.12 & Q & 0.80 & Q & 0.93 \\
515 & AGAL359.887+00.031\_A & 359.89 & 0.03 & Q & 0.59 & Q & 0.83 \\
516 & AGAL359.912-00.129\_S & 359.91 & -0.13 & Q & 0.89 & Q & 0.91 \\
517 & AGAL359.916-00.047\_B & 359.92 & -0.05 & P & 0.70 & P & 0.83 \\
518 & AGAL359.962+00.129\_A & 359.96 & 0.13 & Q & 0.88 & Q & 0.89 \\
519 & *AGAL359.962+00.129\_B & 359.96 & 0.13 & Q & 0.60 & A & 0.52 \\
520 & AGAL359.972-00.072\_S & 359.97 & -0.07 & Q & 0.54 & Q & 0.78 \\
521 & AGAL359.989+00.087\_S & 359.99 & 0.09 & Q & 0.87 & Q & 0.89 \\
522 & AGAL359.990+00.107\_S & 359.99 & 0.11 & Q & 0.95 & Q & 0.92 \\
\end{longtable}
\twocolumn

\bsp	
\label{lastpage}
\end{document}